\documentclass[journal]{IEEEtran}
\usepackage[table]{xcolor}
\usepackage{cite}
\usepackage{graphicx}
\usepackage{lipsum}
\usepackage{stfloats}
\usepackage{bm}
\usepackage{amsmath,amssymb}
\usepackage{amsthm}
\usepackage{pifont}
\usepackage{bbding}
\usepackage{booktabs}
\usepackage{ctable}
\usepackage{threeparttable}
\usepackage{footmisc}
\usepackage{framed} 
\usepackage{array}
\usepackage{multirow}
\usepackage{longtable}
\usepackage{rotating}
\usepackage{fancyhdr}
\usepackage{lastpage}
\usepackage{layout}
\usepackage{wrapfig}
\usepackage{xcolor}
\usepackage{tabularx}
\usepackage{tabu}
\usepackage{enumitem}
\usepackage{algorithm}
\usepackage{algorithmic}
\usepackage{url}
\usepackage{dsfont}
\usepackage{arydshln}

\makeatletter
\newcommand{\removelatexerror}{\let\@latex@error\@gobble}
\makeatother

\makeatletter
\newenvironment{breakablealgorithm}
  {
   \begin{center}
     \refstepcounter{algorithm}
     \hrule height.8pt depth0pt \kern2pt
     \renewcommand{\caption}[2][\relax]{
       {\raggedright\textbf{\ALG@name~\thealgorithm} ##2\par}%
       \ifx\relax##1\relax 
         \addcontentsline{loa}{algorithm}{\protect\numberline{\thealgorithm}##2}%
       \else 
         \addcontentsline{loa}{algorithm}{\protect\numberline{\thealgorithm}##1}%
       \fi
       \kern2pt\hrule\kern2pt
     }
  }{
     \kern2pt\hrule\relax
   \end{center}
  }
\makeatother

\newtheorem{myDef}{\textbf{Definition}}

\newtheorem{myPropos}{\textbf{Proposition}}
\graphicspath{{Figures/}}
\DeclareGraphicsExtensions{.png, .eps}

\begin{document}
	\title{Wireless Resource Optimization in Hybrid Semantic/Bit Communication Networks}
	\author{
	Le Xia,~\IEEEmembership{Member,~IEEE},
				 Yao Sun,~\IEEEmembership{Senior~Member,~IEEE},\\
				 Dusit Niyato,~\IEEEmembership{Fellow,~IEEE},
				 Lan Zhang,~\IEEEmembership{Member,~IEEE},
				 and Muhammad Ali Imran,~\IEEEmembership{Fellow,~IEEE}
	\thanks{
	
	This research of Dusit Niyato was supported by the National Research Foundation, Singapore, and Infocomm Media Development Authority under its Future Communications Research \& Development Programme, Defence Science Organisation (DSO) National Laboratories under the AI Singapore Programme (FCP-NTU-RG-2022-010 and FCP-ASTAR-TG-2022-003), Singapore Ministry of Education (MOE) Tier 1 (RG87/22), and the NTU Centre for Computational Technologies in Finance (NTU-CCTF). The research of Lan Zhang was partially supported by the US National Science Foundation CNS-2418308, CCF-2427316, and CCF-2426318. Preliminary results of this work will be presented in part at the IEEE Global Communications Conference (GlobeCom), 2024~\cite{xia2024hybrid}. (\textit{Corresponding author: Yao Sun}.)
	
	Le Xia, Yao Sun (\textit{Corresponding author: Yao Sun}.), and Muhammad Ali Imran are with the James Watt School of Engineering, University of Glasgow, Glasgow G12 8QQ, UK (e-mail: xiale1995@outlook.com; \{Yao.Sun, Muhammad.Imran\}@glasgow.ac.uk).
	
	Dusit Niyato is with the College of Computing and Data Science, Nanyang Technological University, Singapore 639798 (e-mail: dniyato@ntu.edu.sg).

	Lan Zhang is with the Department of Electrical and Computer Engineering, Clemson University, Clemson, South Carolina 29634, USA (e-mail: lan7@clemson.edu).
	
}
}	
	\maketitle
	\begin{abstract}

	
	Recently, semantic communication (SemCom) has shown great potential in significant resource savings and efficient information exchanges, thus naturally introducing a novel and practical cellular network paradigm where two modes of SemCom and conventional bit communication (BitCom) coexist.
	Nevertheless, the involved wireless resource management becomes rather complicated and challenging, given the unique background knowledge matching and time-consuming semantic coding requirements in SemCom.
	To this end, this paper jointly investigates user association (UA), mode selection (MS), and bandwidth allocation (BA) problems in a hybrid semantic/bit communication network (HSB-Net).
	Concretely, we first identify a unified performance metric of message throughput for both SemCom and BitCom links.
	Next, we specially develop a knowledge matching-aware two-stage tandem packet queuing model and theoretically derive the average packet loss ratio and queuing latency.
	Combined with practical constraints, we then formulate a joint optimization problem for UA, MS, and BA to maximize the overall message throughput of HSB-Net.
	 Afterward, we propose an optimal resource management strategy by utilizing a Lagrange primal-dual transformation method and a preference list-based heuristic algorithm with polynomial-time complexity.
	 Numerical results not only demonstrate the accuracy of our analytical queuing model, but also validate the performance superiority of our proposed strategy compared with different benchmarks.
	
	
	\end{abstract}
	

	\IEEEpeerreviewmaketitle
	
	\section{Introduction}
	\IEEEPARstart{S}{emantic} communication (SemCom) has recently attracted widespread attention as an emerging communication paradigm, promising to significantly alleviate the scarcity of wireless resources in next-generation cellular networks~\cite{zhang2022toward}.
	By embedding cutting-edge sophisticated deep learning (DL) models into wireless terminal devices~\cite{li2023deep}, SemCom is capable of providing mobile users (MUs) with a variety of high-quality, large-capacity, and multimodal services, including typical multimedia content (e.g., text~\cite{xie2021deep}, image~\cite{10122232}, and video streaming~\cite{xia2023wiservr}) and artificial intelligence-generated content (AIGC)~\cite{xia2023generative}.
	Different from the conventional bit communication (BitCom) mode that aims at the precise reception of transmitted bits, SemCom focuses more on the accurate delivery of the true meanings implied in source messages.
	
	Specifically, a DL-based semantic encoder deployed in the transmitter first filters out redundant content from source information to extract core semantics that require fewer bits for transmission.
	After necessary channel encoding and decoding, the core meanings are then accurately restored from the received bits via a jointly trained semantic decoder, even with intolerable bit errors in data propagation~\cite{sun2024s}.
	Notably, either the semantic encoding or decoding is executed based on background knowledge pertinent to the delivered messages, and the higher the knowledge-matching degree between transceivers, the lower the semantic ambiguity in recovered meanings.\footnote{{\label{ft1}}Note that various reasonable assumptions regarding knowledge matching in SemCom can be adopted. If we consider the DL-based training database as knowledge, its matching degree can be the overlap ratio of achievable learning tasks between the transceiver~\cite{10122232}, and if taking into account the knowledge graph, its matching degree can be calculated based on graph similarity~\cite{10614204}.}
	Consequently, it is envisioned that the introduction of SemCom can ensure efficient and reliable information exchanges and save considerable spectrum resources.
	
	Recap the recent advancement of SemCom, there have been some noteworthy related works propelling both its information-theoretic and link-level systematic modelings.
	In~\cite{bao2011towards} and~\cite{basu2014preserving}, Bao~\textit{et al.} quantitatively measured semantic entropy by first proposing a semantic channel-coding theorem, which is based on the information logical probability defined by Carnap and Bar-Hillel in~\cite{carnap1952outline}.
	Besides, Liu~\textit{et al.}~\cite{liu2022indirect} identified the semantic rate-distortion function by leveraging the intrinsic state and extrinsic observation of information in a memoryless source case.
	As for the semantic transceiver design, Xie~\textit{et al.}~\cite{xie2021deep} devised a Transformer-enabled end-to-end SemCom system for reliable text transmission, and then upgraded this system to be lightweight in~\cite{xie2020lite}.
	Moreover, Xia~\textit{et al.}~\cite{xia2023wiservr} presented a mobile virtual reality SemCom framework that can guarantee high-performance semantic extraction and frame recovery for delivering $360^{\circ}$ video streaming.

	In parallel, several other preliminary studies related to SemCom have further investigated the wireless resource management issue from a networking perspective.
	Powered by deep reinforcement learning algorithms, Zhang~\textit{et al.}~\cite{10122232} adopted a dynamic resource allocation scheme to maximize the long-term transmission efficiency in task-oriented SemCom networks.
	In~\cite{10032275}, Yang~\textit{et al.} exploited a probability graph and a rate splitting method to achieve energy-efficient SemCom networks on both transmission and computation.
	Likewise, a quantum key distribution-secured resource management framework was considered by Kaewpuang~\textit{et al.}~\cite{10207065} for the edge devices communicating semantic information.
	Apart from these, Xia~\textit{et al.}~\cite{10261329} specially developed a bit-rate-to-message-rate transformation function along with a new semantic-aware metric called system throughput in message to jointly optimize user association (UA) and bandwidth allocation (BA) problems in SemCom-enabled cellular networks.
	
	
	Nevertheless, notice that the current colossal infrastructures and user groups in BitCom that cannot be completely replaced at one time, while we are embracing the unprecedented potential of SemCom for efficient information exchanges and wireless resource savings.
	Such a trend dooms the future wireless networks toward a more flexible, targeted, and economical fusion architecture.
	Although some related works, e.g.,~\cite{evgenidis2024hybrid}, have explored the point-to-point transmission selection problem between SemCom and BitCom, there is still a lack of relevant investigations on wireless resource management from a networking perspective, i.e.,~\textit{hybrid semantic/bit communication networks} (HSB-Nets), in which both modes are capable of realizing transmission between multiple MUs and BSs.
	Furthermore, the resource optimization in HSB-Nets is expected to yield a host of benefits, such as flexible and targeted service provisioning, adequate resource utilization, and satisfactory user experience on semantic performance.
	
	Nevertheless, since SemCom typically requires more data processing time but produces higher semantic performance than BitCom at each transceiver, choosing a proper mode for each MU should be rather complicated and challenging.
	Most uniquely, the varying degrees of background knowledge matching among MUs can also affect the amount of allocated bandwidth in combination with different channel conditions.
	As such, if aiming at high semantic fidelity and low latency for a large-scale HSB-Net, we are encountering the following three fundamental challenges in resource management:
	\begin{itemize}
		\item \emph{Challenge 1: How to unify performance metrics for both SemCom and BitCom in the HSB-Net?}
		Given the core mechanism of meaning delivery in SemCom, traditional metrics in BitCom, like bit rate or bit throughput, are evidently no longer applicable to the SemCom links.
		Especially in such a hybrid scenario, it becomes necessary to align SemCom and BitCom to the same assessment basis to facilitate subsequent performance comparisons or overall network optimization, which raises the first nontrivial point.
		
		\item \emph{Challenge 2: How to mathematically characterize the unique semantic-coding process in SemCom when combined with bit transmission?}
		Note that SemCom involves an extra semantic-coding process compared with BitCom before the bit data transmission at each link, which can be characterized from a packet-queuing perspective.
		In the semantic-coding process, due to diverse knowledge-matching degrees among different SemCom-enabled MUs, semantic data packet interpretation rates can vary~\cite{xia2023xurllc}, thereby resulting in distinct queuing delay and reliability performance.
		Combined with the subsequent indispensable packet-transmission queuing process, all of these constitute the second difficulty.
		
		\item \emph{Challenge 3: How to determine the best communication mode for each MU with the joint consideration of UA and BA to optimize overall network performance?} 
		Generally, each MU can select only one of the SemCom and BitCom modes at a time during the UA process, subject to its current knowledge-matching degree, channel condition, desired service quality, as well as latency and reliability budgets.
		Such a new mode selection (MS) problem, coupled with inherent practical constraints such as limited bandwidth resources and the single-base station (BS) association requirement, poses the third challenge, i.e., seeking an optimal resource management strategy for the UA, MS, and BA to jointly optimize overall network performance in the HSB-Net.
	\end{itemize}
	
	In response to the challenges outlined above, in this paper, we systematically investigate the UA, MS, and BA problems in the uplink of the HSB-Net and correspondingly propose an optimal strategy with the awareness of unique SemCom characteristics.
	Simulation results not only demonstrate the accuracy of our theoretical analysis for semantic data packet queuing, but also showcase the performance superiority of the proposed resource management solution in terms of realized message throughput compared with four benchmarks.
    Accordingly, our main contributions are summarized as follows:
    \begin{itemize}
		\item
		We unify the performance metrics for both SemCom and BitCom links by introducing the bit-rate-to-message-rate transformation mechanism to measure their respective achievable message throughputs.
		In this regard, the stochasticity of knowledge matching degree and channel state are particularly taking into account over different time slots.
		Correspondingly, we then formulate an optimization problem to maximize the time-averaged overall message throughput of the HSB-Net by jointly correlating the UA, MS, and BA-related indicators.
		These first address the aforementioned~\textit{Challenge 1}.
		\item
		We specially model a two-stage tandem queue for each SemCom-enabled MU to capture the entire queuing process of its locally generated semantic packets, which fully incorporates the semantic coding and knowledge-matching characteristics with the traditional packet transmission.
		On this basis, the steady-state average packet loss ratio and queuing delay in both SemCom and BitCom cases are then mathematically derived to post the reliability and latency requirements in subsequent optimization.
		The contribution directly addresses~\textit{Challenge 2}.
		\item
		We theoretically prove the monotonicity of allocated bandwidth with respect to reliability and latency, and then develop an efficient resource management strategy to jointly solve the UA, MS, and BA problems with polynomial-time complexity.
		Specifically, the minimum bandwidth threshold is first fixed for each SemCom and BitCom link, following by a Lagrange primal-dual method and a preference list-based heuristic algorithm to finalize the UA and MS solutions.
		Afterward, the optimal BA strategy is further obtained by reallocating the remaining bandwidth of each BS to all its associated MUs.
		In this way,~\textit{Challenge 3} is finally well tackled.
	\end{itemize}
    
    The remainder of this paper is organized as follows.
    Section II first introduces the system model of HSB-Net.
    Then, the queuing analysis for both SemCom and BitCom cases are presented, and the corresponding joint resource management problem is formulated in Section III.
    In Section IV, we illustrate the proposed optimal UA, MS, and BA strategy.
    Numerical results are demonstrated and discussed in Section V, followed by the conclusions in Section VI.
    
    \section{System Model}
	
	\subsection{HSB-Net Scenario}
\begin{figure}[t]
		\centering
		\includegraphics[width=0.48\textwidth]{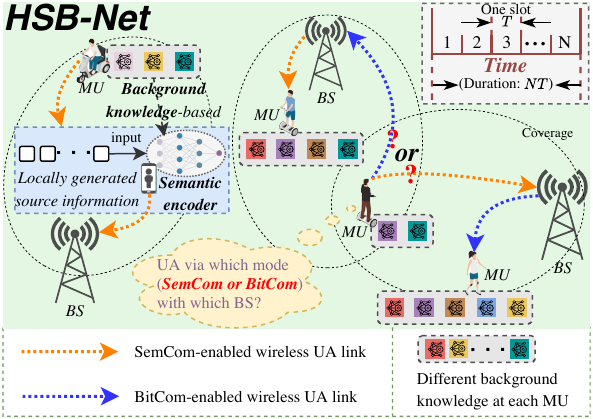} 
		\caption{The HSB-Net scenario involving UA, MS, and BA in one time block.}
		\label{Scenario}
    \end{figure}

	Consider an HSB-Net scenario as depicted in Fig.~\ref{Scenario}, the total of $U$ MUs are distributed within the coverage of $S$ BSs, where two communication modes of SemCom and BitCom are available for all MUs, while each MU can only select one mode and be associated with one BS at a time.
	Herein, let $x_{ij}\in \{0,1\}$ denote the binary UA indicator, where $x_{ij} = 1$ means that MU $i \in \mathcal{U}=\{1,2,\cdots,U\}$ is associated with BS $j \in \mathcal{J}=\{1,2,\cdots,J\}$, and $x_{ij} = 0$ otherwise.
	Besides, we specially define the binary MS indicator as $y_{ij}\in \{0,1\}$, where $y_{ij} = 1$ represents that the SemCom mode is selected for the link between MU $i$ and BS $j$, and $y_{ij} = 0$ indicates that the BitCom mode is selected.\footnote{It is worth pointing out that $y_{ij}$ is applicable to be an effective MS indicator only when $x_{ij}=1, \forall \left( i,j\right) \in\mathcal{U}\times \mathcal{J}$.}
	Meanwhile, the amount of bandwidth resource that BS $j$ assigns to MU $i$ is denoted as $z_{ij}$, while the total bandwidth budget of BS $j$ is denoted as $Z_{j}$.	
	Moreover, time is equally partitioned into $N$ consecutive time slots, each with the same duration length $T$.
	
	\subsection{Network Performance Metric}
	For the wireless propagation model, first let $\gamma_{ij}(t)$ denote the signal-to-interference-plus-noise ratio (SINR) of the link between MU $i$ and BS $j$ at time slot $t,\ t = 1,2,\cdots,N$.\footnote{The SINR calculation can be based on various methods like RSRP~\cite{afroz2015sinr}, which will not affect the remaining modeling and solutions. Due to space limitations, here we directly denote $\gamma_{ij}(t)$ to make room for other modeling.}
	Note that $\gamma_{ij}(t)$ is assumed to be an independent and identically distributed (i.i.d.) random variable for different slots but remain constant during one slot~\cite{guo2019resource,moustakas2013sinr}.
	Since the conveyed message itself becomes the sole focus of precise reception in SemCom rather than traditional transmitted bits in BitCom, we proceed with the performance metric developed in our previous work~\cite{10261329} to measure the message rate for each SemCom-enabled MU via employing a bit-rate-to-message-rate (B2M) transformation function.
	Compared with the conventional bit rate, the message rate is a preferable indicator to concentrate on the true semantics that each MU desires to convey in the source information.
	To be specific, the B2M function is to output the semantic channel capacity (i.e., the achievable message rate in units of messages per unit time, \textit{msg/s}) from input traditional Shannon channel capacity (i.e., the achievable bit rate in units of bits per unit time, \textit{bit/s}) under the discrete memoryless channel in SemCom systems.\footnote{The B2M is actually derived from a semantic information-theoretical perspective by following the work in~\cite{bao2011towards}. Relevant details are already beyond the scope of this paper and thus will not be discussed in-depth.}
	If in an ideal SemCom condition, i.e., the transmitter and receiver have identical semantic reasoning capability and perfectly matching background knowledge, the B2M function can be approximated as linear~\cite{bao2011towards}.
	However, the B2M can also involve stochastic variables in the case of knowledge mismatch, resulting in the presentation of random message rates. 
	Given this, let $\Re_{ij}(\cdot)$ denote the B2M function of the SemCom link between MU $i$ and BS $j$, its instantaneous achievable message rate in time slot $t$ should be
	\begin{equation}
		M_{ij}^{S}(t)=\beta_{i}(t)\Re_{ij}\!\left(z_{ij}\log_{2}\left(1+\gamma_{ij}\left(t\right)\right)\right).\label{SCmessagethroughput}
	\end{equation}
	Here, $\beta_{i}(t)$ represents the knowledge-matching degree between MU $i$ and its communication counterpart at slot $t$, which is an i.i.d. random Gaussian variable ranging from $0$ to $1$~\cite{10261329}, having mean $\tau_i$.
	To provide more details here, each message is first assumed to be associated with a specific SemCom service type based on Footnote~\ref{ft1}. Then, compared with the perfectly knowledge-matching case, only the messages related to the overlapped services can be effectively encoded/decoded in the knowledge-mismatching state in each slot, and $\beta_{i}(t)$ is the overlap proportion.
	Combined with the fact that the generation of source messages is generally a stochastic process~\cite{liu2022indirect}, therefore, $\beta_{i}(t)$ is deemed as a random variable.
	In addition, other factors like channel encoding scheme that may affect the message-rate measurement are assumed to be identical between different SemCom-enabled MUs for simplicity.
	
	Likewise, for the BitCom link between MU $i$ and BS $j$, considering it has an average B2M transformation ratio,\footnote{This assumption is justified since the source-and-channel coding of BitCom for source information typically follows prescribed codebooks, and the variable length coding is adopted~\cite{merkle1978secure,xia2023generative}. Hence, based on each link's known channel state information, the proportion of messages that can be effectively decoded from a certain amount of transmitted bits can be averaged.} denoted by $\rho_{ij}$, to align with the semantic performance measurement of SemCom.
	In other words, we assume that each message in BitCom can be encoded into bits of fixed length on average~\cite{gao2024importance}, i.e., the reciprocal of $\rho_{ij}$, and thus its instantaneous achievable message rate in slot $t$ is given by
	\begin{equation}
		\begin{aligned}
			M_{ij}^{B}(t)=\rho_{ij}z_{ij}\log_{2}\left( 1+\gamma_{ij}\left(t\right)\right),\ 0<\rho_{ij}<1.\label{BCmessagethroughput}
		\end{aligned}
	\end{equation}
	
	As such, if taking into account both SemCom (i.e., $y_{ij}=1$) and BitCom (i.e., $y_{ij}=0$) cases, we obtain the time-averaged message rate of each link as
	\begin{equation}
		\label{overallthroughput}
			M_{ij}=\frac{1}{N}\sum_{t=1}^{N}\left[y_{ij}M_{ij}^{S}(t)+(1-y_{ij})M_{ij}^{B}(t)\right].
	\end{equation}
	
	\subsection{Queuing Model}
	\begin{figure}[t]
		\centering
		\includegraphics[width=0.48\textwidth]{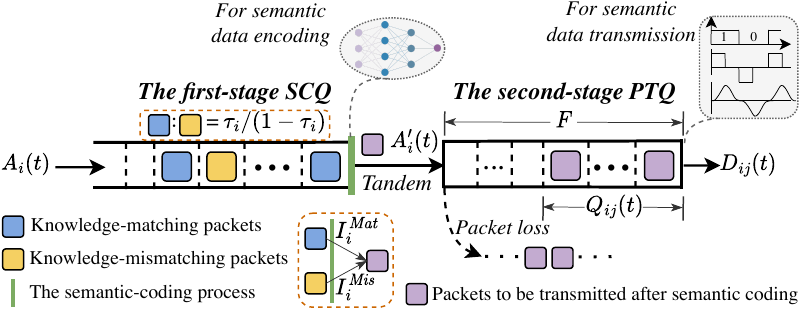} 
		\caption{The two-stage tandem queue model at each SemCom-enabled MU.}
		\label{Queuing}
    \end{figure}
	In this work, we focus on the differences in queuing models between SemCom and BitCom during data uplink transmission, where the queuing delay is employed as the latency metric to characterize the average sojourn time of a data packet in the queue buffer at each MU in the HSB-Net.
	Besides only considering packet transmission queuing in BitCom like many existing studies~\cite{meshkati2009energy,xu2018energy,ding2022two,wang2019effective}, the queuing delay should take into account the semantic-coding process newly introduced in SemCom due to the intrinsic limited computation capability of each MU.
	Given the illustration in Fig.~\ref{Queuing}, we first provide the following definition for clarity.
	\begin{myDef}
		\label{Def1}
		A SemCom-enabled MU has a two-stage tandem queue,\footnote{A two-stage tandem queue implies that the output of the first queue becomes the input of the second, and the packet processing in the two queues is independent of each other~\cite{wang2019effective,wu2006dynamic}.} named Semantic-Coding Queue (SCQ) and Packet-Transmission Queue (PTQ). As for a BitCom-enabled MU, only one PTQ model is considered for its data packet uplink transmission without the involvement of SCQ.
	\end{myDef}
	To preserve the generality, the SCQ is assumed with infinite-size memory to handle all locally generated SemCom services, while the PTQ has a finite-size buffer that can accommodate up to $F$ data packets to align with practical resource limitations and scheduling.\footnote{Note that the SCQ can also be modeled with a finite-size buffer whose queuing latency is derived similarly to that of the PTQ. Likewise, the above rationale applies to the PTQ as well.}
	Moreover, the packets in both SCQ and PTQ are queued in a first-come-first-serve manner.
	
	Based on the above, if a Poisson arrival process with average rate $\lambda_{i}$ (in \textit{packets/s}) of initial data packet generation is assumed for each MU $i$ ($\forall i \!\in \!\mathcal{U}$), the number of arrival packets during slot $t$, denoted as $A_{i}(t)$, has the probability mass function (PMF) as follows:
	\begin{equation}
		\label{initialarrival}
			\Pr\left\{A_{i}\left(t\right)=k\right\}=\frac{\left(\lambda_{i}T\right)^{k}}{k!}\exp{\left(-\lambda_{i}T\right)},\ k=0, 1, 2, \cdots.
	\end{equation}
	
	For the PTQ in both SemCom and BitCom cases, its packet departure rate depends on the number of packets sent out from MU $i$ to BS $j$ ($\forall j \!\in \!\mathcal{J}$) during slot $t$, denoted as $D_{ij}\left(t\right)$, which has the PMF as
	\begin{equation}
		\label{PTQDepart}
		\begin{aligned}
			\Pr&\left\{D_{ij}\left(t\right)=k\right\}=\Pr\left\{\left\lfloor \frac{Tz_{ij}\log_{2}\left( 1+\gamma_{ij}\left(t\right)\right)}{L}\right\rfloor =k\right\}\\
			&=\Pr\left\{\gamma_{ij}\left(t\right)\leqslant 2^{\frac{(k+1)L}{Tz_{ij}}}\!-\!1\right\}\!-\!\Pr\left\{\gamma_{ij}\left(t\right)\leqslant 2^{\frac{kL}{Tz_{ij}}}\!-\!1\right\}.
		\end{aligned}
	\end{equation}
	Here, $\left\lfloor\cdot\right\rfloor$ is the floor function that outputs the largest integer less than or equal to the input value, and all packets have the same size of $L$ bits, $k=0, 1, 2, \cdots$.
	Clearly, given any reasonable probability distribution approximation of the SINR $\gamma_{ij}\left(t\right)$ (e.g., Gaussian distribution~\cite{moustakas2013sinr} or generalized Gamma distribution~\cite{li2005distribution}), applying its cumulative distribution function (CDF) directly yields the close-form expression of \eqref{PTQDepart}.
	Besides, it is noteworthy that the obtained PMF of $D_{ij}\left(t\right)$ should be independent of time slot index $t$, as the randomness of each physical link's SINR is generally $t$-independent~\cite{guo2019resource}.

	Next, we model the packet departure of SemCom-enabled SCQ and the packet arrival of SemCom-enabled PTQ, respectively.
	As mentioned earlier, each data packet generated at a SemCom-enabled sender MU requires a certain type of background knowledge, resulting in either a knowledge-matching or knowledge-mismatching state with its receiver.
	For illustration, let $I_{i}^{\mathit{Mat}}$ denote the semantic-coding time required by a knowledge-matching packet with mean $1/\mu_{i}^{\mathit{Mat}}$ (in \textit{s/packet}), and let $I_{i}^{\mathit{Mis}}$ denote the semantic-coding time required by a knowledge-mismatching packet with mean $1/\mu_{i}^{\mathit{Mis}}$ ($\mu_{i}^{\mathit{Mat}}\!>\!\mu_{i}^{\mathit{Mis}}$ in practice\footnote{Note that the content in knowledge-mismatching packets require more computation resources on further model fine-tuning or knowledge-sharing to realize the same-level accurate contextual reasoning and interpretation as knowledge-matching ones~\cite{luo2022semantic}, thus leading to additional processing time.}).
	Without loss of generality, $I_{i}^{\mathit{Mat}}$ and $I_{i}^{\mathit{Mis}}$ are assumed to be two exponential random variables independent of each other, which are determined by the specific semantic computing capability available at the MU $i$'s terminal device.
	Having these, it is seen that the overall service time distribution of packets at SCQ should be treated as a general distribution~\cite{xia2023xurllc}.
	Let us denote the average packet queuing latency of SCQ at each SemCom-enabled MU $i$ by $\delta_{i}^{\mathit{S_1}}$, which will be analyzed in detail in the next section.
	
	As for the number of packets arriving at the SemCom-enabled PTQ in slot $t$, denoted by $A_{i}'(t)$, it should exactly be the number of packets leaving its tandem SCQ in the same slot, according to the two-stage tandem structure in Definition~\ref{Def1}.
	Meanwhile, due to the exponential departure assumptions, the knowledge-matching packets leaving the SCQ follow a Poisson distribution with mean $\mu_{i}^{\mathit{Mat}}$, while the knowledge-mismatching packets leave as a Poisson process with mean $\mu_{i}^{\mathit{Mis}}$~\cite{ross2014introduction}.
	The former event occurs with probability $\tau_{i}$ and the latter happens with probability $(1-\tau_{i})$.
	As such, $A_{i}'(t)$ should still satisfy the Poisson distribution with a PMF of
	\begin{equation}
		\begin{aligned}
		\label{PTQmixedarrival}
		\Pr\left\{A'_{i}\!\left(t\right)=k\right\}=&\frac{\left(\lambda_{i}'T\right)^{k}}{k!}\exp{\left(-\lambda_{i}'T\right)},\ k=0, 1, 2, \cdots,
		\end{aligned}
	\end{equation}
	where $\lambda_{i}'$ is the average arrival rate (in \textit{packets/s}), given as
	\begin{equation}
		\label{SPTQmean}
		\lambda_{i}'=\tau_{i}\mu_{i}^{\mathit{Mat}}+(1-\tau_{i})\mu_{i}^{\mathit{Mis}}.
	\end{equation}
	
	
	Considering the limited buffer size $F$ of PTQ, we further assume that in any $t$, the packets to be transmitted leave the queue first and then the arriving packets enter it.
	Hence, the evolution of its queue length between two consecutive slots is
	\begin{equation}
		\label{PTQevolution}
		Q_{ij}\left(t+1\right)\triangleq \min\left\{\max\left\{Q_{ij}\left(t\right)\!-\!D_{ij}\left(t\right),0\right\}\!+\!A'_{i}\!\left(t\right),F\right\},
	\end{equation}
	where $Q_{ij}\left(t\right)$ denotes the queue length of PTQ for the link between MU $i$ and BS $j$ at slot $t$, $t=1,2,\cdots,N-1$.
	
	Since the focus is on the semantic packet queuing process, the packet queuing loss and queuing delay need to be carefully analyzed. Note that those packets arriving at a fully-loaded PTQ in each slot will be blocked and dropped, which can affect achievable communication reliability and message rate performance.
	Accordingly, let $\theta_{ij}^{\mathit{S}}$ and $\theta_{ij}^{\mathit{B}}$ denote the average packet loss ratio of SemCom-enabled PTQ and BitCom-enabled PTQ, respectively, and each represents the proportion of packets failed to be delivered to all arriving packets.
	Likewise, let $\delta_{ij}^{\mathit{S_2}}$ and $\delta_{ij}^{\mathit{B}}$ denote the average packet queuing latency of SemCom-enabled PTQ and BitCom-enabled PTQ, respectively.
	Combined with the $\delta_{i}^{\mathit{S_1}}$ defined before, we obtain the overall average queuing latency of the link between SemCom-enabled MU $i$ and BS $j$ as $\delta_{ij}^{S}=\delta_{i}^{\mathit{S_1}}+\delta_{ij}^{\mathit{S_2}}$.
	
	When considering both SemCom (i.e., $y_{ij}=1$) and BitCom (i.e., $y_{ij}=0$), the average queuing latency experienced by the link between any MU $i$ and BS $j$ should be
	\begin{equation}
		\label{overallqueuinglatency}
			\delta_{ij}=y_{ij}\delta_{ij}^{S}+(1-y_{ij})\delta_{ij}^{B}.
	\end{equation}
 	Similarly, the average packet loss ratio that indicates the communication reliability of the link is found by
 	\begin{equation}
 		\label{overalllossratio}
 			\theta_{ij}=y_{ij}\theta_{ij}^{S}+(1-y_{ij})\theta_{ij}^{B}.
 	\end{equation}
 	
 	In the subsequent section, we elaborate the derivations for the mathematical expressions of $\delta_{i}^{\mathit{S_1}}$, $\delta_{ij}^{\mathit{S_2}}$, and $\theta_{ij}^{\mathit{S}}$.
 	Recalling the BitCom-enabled PTQ model and the SemCom-enabled PTQ model, it is seen that their sole distinction lies in their packet arrival processes, in which the former follows~\eqref{initialarrival} and the latter follows~\eqref{PTQmixedarrival}.
 	Therefore, $\delta_{ij}^{\mathit{B}}$ and $\theta_{ij}^{\mathit{B}}$ can be easily derived using the similar procedure as for $\delta_{ij}^{\mathit{S_2}}$ and $\theta_{ij}^{\mathit{S}}$.
 	
 	\section{Queuing Analysis and Problem Formulation}
 	\subsection{Queuing Analysis for SCQ and PTQ}
 	First for the SemCom-enabled SCQ, it should be noted that the average proportion of knowledge-matching packets to the total number of packets in the queue is exactly equal to the average knowledge-matching degree $\tau_{i}$ between MU $i$ and its receiver.\footnote{This observation holds true when examined on a large timescale, and it assumes that each packet has the same probability of being generated locally.}
 	Combined with the general distribution conclusion obtained earlier, the average semantic-coding time of a packet in the SCQ, denoted by $I_{i}$, becomes $I_{i}=\tau_{i}I_{i}^{\mathit{Mat}}+\left(1-\tau_{i}\right)I_{i}^{\mathit{Mis}}$.
	Since $I_{i}^{\mathit{Mat}}$ and $I_{i}^{\mathit{Mis}}$ are independent of each other, we have its expectation as $\mathds{E}\left[I_{i}\right]=\tau_{i}/\mu_{i}^{\mathit{Mat}}+\left(1-\tau_{i}\right)/\mu_{i}^{\mathit{Mis}}$ and its variance as $\mathds{V}\left(I_{i}\right)=\left(\tau_{i}/\mu_{i}^{\mathit{Mat}}\right)^2+\left(\left(1-\tau_{i}\right)/\mu_{i}^{\mathit{Mis}}\right)^2$.
	Owing to the Markovian packet arrival and general-distribution packet departure patterns, the SCQ can be modeled as an M/G/1 system, which has been widely used to capture data traffic in wireless networks~\cite{xia2023xurllc,ding2022two}.
	In this case, we can directly apply the~\textit{Pollaczek-Khintchine formula}~\cite{pollaczek1930aufgabe} to calculate the steady-state average packet queuing latency of SCQ $\delta_{i}^{\mathit{S_1}}$, which is expressed in \eqref{SCQ_queuing_delay} as shown at the bottom of the next page.\footnote{Applying the Pollaczek-Khintchine formula implies a prerequisite that $\lambda_{i}\mathds{E}\left[I_{i}\right]\!< \!1$ must be satisﬁed to guarantee a steady-state M/G/1 system~\cite{ross2014introduction}. Therefore, we consider that in the SCQ, the packet departure rate exceeds the packet arrival rate to make its queuing latency finite and solvable.}
	Further noting that either $I_{i}^{\mathit{Mat}}$ or $I_{i}^{\mathit{Mis}}$ in \eqref{SCQ_queuing_delay} is independent of time slot index $t$, thus $\delta_{i}^{\mathit{S_1}}$ should be deemed a constant.
	\begin{figure*}[hb]
		\centering
		\hrulefill
		\begin{equation}
		\label{SCQ_queuing_delay}
			\begin{aligned}
				\delta_{i}^{\mathit{S_1}}&=\frac{\lambda_{i}\left(\mathds{E}^{2}\left[I_{i}\right]+\mathds{V}\left(I_{i}\right)\right)}{2\left(1-\lambda_{i}\mathds{E}\left[I_{i}\right]\right)}+\mathds{E}\left[I_{i}\right]= \frac{\lambda_{i}\left[\tau_{i}\left(1-\tau_{i}\right)/\mu_{i}^{\mathit{Mat}}\mu_{i}^{\mathit{Mis}}+\left(\tau_{i}/\mu_{i}^{\mathit{Mat}}\right)^2+\left(\left(1-\tau_{i}\right)/\mu_{i}^{\mathit{Mis}}\right)^2\right]}{1-\lambda_{i}\tau_{i}/\mu_{i}^{\mathit{Mat}}-\lambda_{i}\left(1-\tau_{i}\right)/\mu_{i}^{\mathit{Mis}}}+\frac{\tau_{i}}{\mu_{i}^{\mathit{Mat}}}+\frac{1-\tau_{i}}{\mu_{i}^{\mathit{Mis}}}.
			\end{aligned}
		\end{equation}
	\end{figure*}
	
	When it comes to the SemCom-enabled PTQ, we first introduce the following proposition to characterize its steady-state queue length $Q_{ij}(t)=0, 1, 2, \cdots, k, \cdots, F$ in slot $t$.
	\begin{myPropos}
	\label{Prop1}
		For each $Q_{ij}(t)$ of PTQ, it must have a solvable and unique steady-state probability vector, denoted as $\bm{\alpha}_{ij}=\left[\alpha_{ij}^{0},\alpha_{ij}^{1},\cdots,\alpha_{ij}^{F}\right]^{T}$, where $\alpha_{ij}^k$ represents the steady-state probability of $Q_{ij}\left(t\right)=k$ when $t$ tends to infinity.
	\end{myPropos}
	\begin{IEEEproof}
		Please see Appendix A.
	\end{IEEEproof}
	
	From Proposition~\ref{Prop1}, the long-term average queue length of $Q_{ij}(t)$ can be obtained by computing its expectation, i.e., $\mathds{E}\left[Q_{ij}(t)\right]=\sum_{k=0}^{F}k\alpha_{ij}^{k}$.
	Moreover, by combining $\bm{\alpha}_{ij}$ with the PMFs of PTQ's packet arrival as in~(\ref{PTQmixedarrival}) and packet departure as in~(\ref{PTQDepart}), the average number of packets dropped at the steady-state PTQ during any slot $t$, denoted by $G_{ij}$, can be calculated by \eqref{Drop} at the bottom of the next page.
	\begin{figure*}[hb]
		\centering
		\hrulefill
		\begin{equation}
			\label{Drop}
			\begin{aligned}
				G_{ij}&=\sum_{l=1}^{F}\alpha_{ij}^{l}\!\left[\sum_{k=0}^{l-1}\Pr\left\{D_{ij}\!=\!k\right\}\!\!\left(\sum_{f=F-l+k}^{\infty}\!\left(f\!+\!l\!-\!k\!-\!F\right)\Pr\!\left\{\!A'_{i}\!=\!f\!\right\}\!\right)\!+\!\left(\sum_{k=l}^{\infty}\Pr\left\{\!D_{ij}\!=\!k\!\right\}\!\right)\!\!\left(\sum_{f=F+1}^{\infty}\!\left(f\!-\!F\right)\Pr\!\left\{\!A_{i}'=f\!\right\}\!\right)\right]\\
				&\qquad +\alpha_{ij}^{0}\!\sum_{k=F+1}^{\infty}(k-F)\Pr\left\{A'_{i}=k\right\}.
			\end{aligned}
		\end{equation}
	\end{figure*}
	As its average total packet arrival rate is $\lambda_{i}'$, we have the steady-state average packet loss ratio of SemCom-enabled PTQ as follows:
	\begin{equation}
	\label{SemComplr}
		\theta_{ij}^{\mathit{S}}=\frac{G_{ij}}{\lambda_{i}'T}=\frac{G_{ij}}{\tau_{i}\mu_{i}^{\mathit{Mat}}T+\mu_{i}^{\mathit{Mis}}T-\tau_{i}\mu_{i}^{\mathit{Mis}}T}.
	\end{equation}
	Hence, the average effective packet arrival rate becomes $\lambda_{i}^{\mathit{eff}}=\left(1-\theta_{ij}^{\mathit{S}}\right)\lambda_{i}'=\tau_{i}\mu_{i}^{\mathit{Mat}}+(1-\tau_{i})\mu_{i}^{\mathit{Mis}}-G_{ij}/T$.
	As such, we can apply Little's law~\cite{little2008little} to finalize the steady-state average queuing latency of SemCom-enabled PTQ as
	\begin{equation}
		\label{PTQ_queuing_delay}
		\begin{aligned}
		\delta_{ij}^{\mathit{S_2}}=\frac{\mathds{E}\left[Q_{ij}(t)\right]}{\lambda_{i}^{\mathit{eff}}}=\frac{\sum_{k=0}^{F}k\alpha_{ij}^{k}}{\tau_{i}\mu_{i}^{\mathit{Mat}}+(1-\tau_{i})\mu_{i}^{\mathit{Mis}}-G_{ij}/T}.
		\end{aligned}
	\end{equation}
	
	Furthermore, to determine the expressions of BitCom-enabled average packet queuing latency $\delta_{ij}^{\mathit{B}}$ and the BitCom-enabled average packet loss ratio $\theta_{ij}^{\mathit{B}}$, the same mathematical methods as the above can be employed, where only the PMF and mean of $A'_{i}(t)$ as in~(\ref{PTQmixedarrival}) in each relevant term need to be substituted with that of $A_{i}(t)$ as in~(\ref{initialarrival}).
	For brevity, the derivation details for $\delta_{ij}^{\mathit{B}}$ and $\theta_{ij}^{\mathit{B}}$ are omitted here.
 	
	\subsection{Problem Formulation}
	For ease of illustration, we first define three variable sets $\bm{x}=\left\{x_{ij}\mid i \in \mathcal{U},j \in \mathcal{J}\right\}$, $\bm{y}=\left\{y_{ij}\mid i \in \mathcal{U},j \in \mathcal{J}\right\}$, and $\bm{z}=\left\{z_{ij}\mid i \in \mathcal{U}, j \in \mathcal{J}\right\}$ that consist of all possible indicators pertinent to UA, MS, and BA, respectively.
	Without loss of generality, the objective is to maximize the overall message throughput (i.e., the sum of the achievable message rates of all MUs) of the HSB-Net by jointly optimizing $(\bm{x},\bm{y}, \bm{z})$, while subject to SemCom-relevant latency and reliability requirements alongside several practical system constraints.
	Notice that the message throughput performance $M_{ij}$ in~\eqref{overallthroughput} is actually the ergodic capacity of each link over the timescale of a block when $N$ is large enough, and thus can be computed through averaging the two time-dependent parameters $\gamma_{ij}(t)$ and $\beta_{i}(t)$ within it~\cite{liang2017resource}.
	Accordingly, if denoting the long-term average of $M_{ij}^{S}(t)$ and $M_{ij}^{B}(t)$ as $\overline{M}_{ij}^{S}$ and $\overline{M}_{ij}^{B}$, respectively, when $N$ tends to infinity in~\eqref{overallthroughput}, our optimization objective becomes
	\begin{equation}
		\label{meanmessagethroughput}
		\begin{aligned}
			&\overline{M}_{ij} = y_{ij}\overline{M}_{ij}^{S}+(1-y_{ij})\overline{M}_{ij}^{B}\\
			&\ =\!y_{ij}\tau_{i}\Re_{ij}\!\left(\!z_{ij}\log_{2}\!\left(1\!+\!\overline{\gamma}_{ij}\right)\!\right)\!+\!\rho_{ij}z_{ij}(1\!-\!y_{ij})\log_{2}\!\left(1\!+\!\overline{\gamma}_{ij}\right)\!,
		\end{aligned}
	\end{equation}
	where $\overline{\gamma}_{ij}$ denotes the mean of $\gamma_{ij}(t)$ and $\tau_{i}$ is the mean of $\beta_{i}(t)$.
	Recalling the average queuing latency $\delta_{ij}$ as in~(\ref{overallqueuinglatency}) and the average packet loss ratio $\theta_{ij}$ as in~(\ref{overalllossratio}), our joint optimization problem $\mathbf{P1}$ is now formulated as follows:
	\begin{align}
	\mathbf{P1}:\ \max_{\bm{x},\bm{y},\bm{z}} \quad & \sum_{i\in \mathcal{U}}\sum_{j\in \mathcal{J}}x_{ij}\overline{M}_{ij}~\label{P1}\\
	{\rm s.t.} \quad & \sum_{j\in \mathcal{J}} x_{ij}= 1,\ \forall i\in \mathcal{U},\tag{\ref{P1}a}\\
	& \sum_{i\in \mathcal{U}}x_{ij} z_{ij}\leqslant Z_{j},\ \forall j\in \mathcal{J},\tag{\ref{P1}b}\\
	& x_{ij}\delta_{ij}\leqslant \delta_{0},\ \forall \left( i,j\right) \in\mathcal{U}\times \mathcal{J},\tag{\ref{P1}c}\\
	& x_{ij}\theta_{ij}\leqslant \theta_{0},\ \forall \left( i,j\right) \in\mathcal{U}\times \mathcal{J},\tag{\ref{P1}d}\\
	& \sum_{j\in \mathcal{J}}x_{ij}\overline{M}_{ij}\geqslant \mathit{M}_{i}^{o},\ \forall i\in \mathcal{U},\tag{\ref{P1}e}\\
	& x_{ij}\!\in \!\left\{ 0,1\right\},\ y_{ij}\!\in \!\left\{ 0,1\right\},\ \forall \left( i,j\right) \in\mathcal{U}\times \mathcal{J}.\tag{\ref{P1}f}
	\end{align}
	Constraints (\ref{P1}a) and (\ref{P1}b) mathematically model the single-BS constraint for UA and the maximum bandwidth resource constraint for BA, respectively.
	Constraints (\ref{P1}c) and (\ref{P1}d) ensure that the average queuing latency and the average packet loss ratio of the link between each MU and its associated BS cannot exceed their respective requirements $\delta_{0}$ and $\theta_{0}$.
	$M_{i}^{th}$ in constraint (\ref{P1}e) represents a minimum message rate threshold for each MU $i$'s association link, while constraint (\ref{P1}f) characterizes the binary properties of both $\bm{x}$ and $\bm{y}$.
	
	Carefully examining $\mathbf{P1}$, it can be observed that the optimization is rather challenging due to several inevitable mathematical obstacles.
	First of all, $\mathbf{P1}$ is clearly an NP-hard problem involving two complicated constraints (\ref{P1}c) and (\ref{P1}d), which leads to a high-complexity solution procedure.
	Another nontrivial point originates from the three different optimization variables, including two integer variables (i.e., $\bm{x}$ and $\bm{y}$) and one continuous variable (i.e., $\bm{z}$).
	In this respect, although we could first relax $\bm{x}$ and $\bm{y}$ to the continuous ones in a conventional manner, the problem after slack should still be nonconvex and the subsequent integer recovery may lead to severe performance compromise~\cite{burer2012non}.
	In full view of the above difficulties, we propose an efficient solution in the next section to solve $\mathbf{P1}$ and obtain the joint optimal strategy for the UA, MS, and BA in the HSB-Net.

	\section{Proposed Resource Management for HSB-Nets}
	To make P1 tractable, each $z_{ij}$ ($\forall (i,j) \in \mathcal{U}\times \mathcal{J}$) is first fixed to two thresholds based on both the SemCom case and the BitCom case, respectively.
	Then, we determine the UA and MS strategies by employing a Lagrange primal-dual method and a devised preference list-based heuristic algorithm.
	On this basis, the BA strategy is then optimally finalized by reallocating the bandwidth of each BS to all its associated MUs while accommodating their respective identified communication modes.
	Finally, we summarize the algorithm of our proposed solution and present its computational complexity analysis.
	
	\subsection{Strategy Determination for UA and MS}
	To make $\mathbf{P1}$ tractable, we first fix $\bm{z}$ to concentrate upon solving $\bm{x}$ and $\bm{y}$.
	Based on the boundary cases of constraints (\ref{P1}c)-(\ref{P1}e), there must be a minimum bandwidth threshold for each SemCom link and BitCom link to simultaneously meet the preset latency, reliability, and message throughout requirements.
	The feasibility behind this approach is established in accordance with the following proposition.
	\begin{myPropos}
	\label{Prop2}
		The steady-state average packet queuing latency $\delta_{ij}$ and average packet loss ratio $\theta_{ij}$ are monotonically non-increasing w.r.t. $z_{ij}$ given any value of $y_{ij}$.
	\end{myPropos}
	\begin{IEEEproof}
		Please see Appendix B.
	\end{IEEEproof}
	Proceeding as in~\cite{10261329}, $\Re_{ij}(\cdot)$ is known to be a monotonically increasing function of $z_{ij}$, and thus $\overline{M}_{ij}$ should also monotonically increase w.r.t. $z_{ij}$ in either the case of $y_{ij}=0$ or $y_{ij}=1$.
	Accordingly, we first consider the boundary situation of the inequality constraint (\ref{P1}e), i.e., $\overline{M}_{ij}=\mathit{M}_{i}^{o}$, the minimum $z_{ij}$ required by the association link between MU $i$ and BS $j$ to perform SemCom (denoted by $z_{ij}^{S\!_{M}}$) and BitCom (denoted by $z_{ij}^{B\!_{M}}$), respectively, can be
	\begin{equation}
		\label{gestsa}
		\begin{aligned}
			z_{ij}^{S\!_{M}}=\frac{\Re_{ij}^{-1}\left(\mathit{M}_{i}^{o}/\tau_{i}\right)}{\log_{2}\left(1+\overline{\gamma}_{ij}\right)}\quad \text{and} \quad z_{ij}^{B\!_{M}}= \frac{\mathit{M}_{i}^{o}}{\rho_{ij}\log_{2}\left(1+\overline{\gamma}_{ij}\right)},
		\end{aligned}
	\end{equation}
	where $\Re_{ij}^{-1}(\cdot)$ indicates the inverse function of $\Re_{ij}(\cdot)$ w.r.t. $z_{ij}$.
	Likewise, in the context of \eqref{overallqueuinglatency} and \eqref{overalllossratio}, we can also obtain the constraint (\ref{P1}c)-based minimum $z_{ij}$ (denoted by $z_{ij}^{S_{\delta}}$ and $z_{ij}^{B_{\delta}}$) and constraint (\ref{P1}d)-based minimum $z_{ij}$ (denoted by $z_{ij}^{S_{\theta}}$ and $z_{ij}^{B_{\theta}}$) in their respective inequality boundary situations.
	It is worth pointing out here that the feasible $z_{ij}^{S_{\delta}}$ solution may not exist if $\delta_{i}^{\mathit{S_1}}>\delta_{0}$, while $\delta_{i}^{\mathit{S_2}}$ cannot be negative.
	In such a case, we set $z_{ij}^{S_{\delta}}=+\infty$ to avoid the possibility of the MU selecting the SemCom mode in the subsequent solution.
	
	Afterward, our aim is to find the optimal $\bm{x}^{*}=\left\{x_{ij}^{*}\mid i \in \mathcal{U}, j\in \mathcal{J}\right\}$ and the optimal $\bm{y}^{*}=\left\{y_{ij}^{*}\mid i \in \mathcal{U}, j\in \mathcal{J}\right\}$ by fixing each SemCom-associated $z_{ij}$ term as $z_{ij}^{S_{th}}\!=\!\max\!\left\{\!z_{ij}^{S\!_{M}}\!,z_{ij}^{S_{\delta}}\!,z_{ij}^{S_{\theta}}\!\right\}$ and each BitCom-associated $z_{ij}$ as $z_{ij}^{B_{th}}\!=\!\max\!\left\{\!z_{ij}^{B\!_{M}}\!,z_{ij}^{B_{\delta}}\!,z_{ij}^{B_{\theta}}\!\right\}$.
	As such, constraints (\ref{P1}c)-(\ref{P1}e) in the primal problem $\mathbf{P1}$ can be all removed, and then $\mathbf{P1}$ degenerates into
	\begin{align}
		\mathbf{P1.1}:\ \max_{\bm{x},\bm{y}} \quad & \sum_{i\in \mathcal{U}}\sum_{j\in \mathcal{J}}x_{ij}\!\left[y_{ij}\overline{M}_{ij}^{S_{th}}+\left(1-y_{ij}\right)\overline{M}_{ij}^{B_{th}}\!\right]~\label{P1.1}\\
		{\rm s.t.} \quad & \sum_{i\in \mathcal{U}}\!x_{ij}\!\left[y_{ij}z_{ij}^{S_{th}}\!+\!\left(1\!-\!y_{ij}\right)z_{ij}^{B_{th}}\!\right]\!\leqslant \!Z_{j},\ \forall j\!\in \!\mathcal{J},\tag{\ref{P1.1}a}\\
		& \text{(\ref{P1}a)},\ \text{(\ref{P1}f)},\tag{\ref{P1.1}b}
	\end{align}
	where let $\overline{M}_{ij}^{S_{th}}\!=\! \tau_{i}\Re_{ij}\!\left(\!z_{ij}^{S_{th}}\!\log_{2}\!\left(1+\overline{\gamma}_{ij}\right)\!\right)$ and $\overline{M}_{ij}^{B_{th}}\!\!=\!\!\rho_{ij}z_{ij}^{B_{th}}\!\log_{2}\!\left( 1+\overline{\gamma}_{ij}\right)$, both are regarded as known constants.
	
	Regarding $\mathbf{P1.1}$, we incorporate constraint (\ref{P1.1}a) into its objective function (\ref{P1.1}) by associating Lagrange multipliers $\bm{\eta}=\left\{\eta_{j}\mid j \in \mathcal{J}\right\}$.
	The associated Lagrange function is presented in \eqref{Lagrange} at the bottom of the next page, in which $\widetilde{L}_{\bm{\eta}}\left(\bm{x},\bm{y}\right)$ is defined for expression brevity.
	\begin{figure*}[hb]
		\centering
		\hrulefill
		\begin{equation}
			\label{Lagrange}
			\begin{aligned}
			L\left(\bm{x},\bm{y},\bm{\eta}\right)&=\sum_{i\in \mathcal{U}}\sum_{j\in \mathcal{J}}x_{ij}\left[y_{ij}\overline{M}_{ij}^{S_{th}}+\left(1-y_{ij}\right)\overline{M}_{ij}^{B_{th}}\right]+\sum_{j\in \mathcal{J}}\eta_{j}\left(Z_{j}-\sum_{i\in \mathcal{U}}x_{ij}\left[y_{ij}z_{ij}^{S_{th}}+\left(1-y_{ij}\right)z_{ij}^{B_{th}}\right]\right)\\
			&=\sum_{i\in \mathcal{U}}\sum_{j\in \mathcal{J}}\left[x_{ij}y_{ij}\left(\overline{M}_{ij}^{S_{th}}-\eta_{j}z_{ij}^{S_{th}}\right)+x_{ij}\left(1-y_{ij}\right)\left(\overline{M}_{ij}^{B_{th}}-\eta_{j}z_{ij}^{B_{th}}\right)\right]+\sum_{j\in \mathcal{J}}\eta_{j}Z_{j}\\
			&\triangleq \widetilde{L}_{\bm{\eta}}\left(\bm{x},\bm{y}\right)+\sum_{j\in \mathcal{J}}\eta_{j}Z_{j}.
			\end{aligned}
		\end{equation}
	\end{figure*}
	That way, the Lagrange dual problem of $\mathbf{P1.1}$ becomes
	\begin{align}
			\mathbf{D1.1}:\ \min_{\bm{\eta}} \quad & H\left(\bm{\eta}\right)=g_{\bm{x},\bm{y}}\left(\bm{\eta}\right)+\sum_{j\in \mathcal{J}}\eta_{j}Z_{j}~\label{D}\\
			{\rm s.t.} \quad & \eta_{j}\geqslant 0,\ \forall j \in \mathcal{J},\tag{\ref{D}a}
	\end{align}
	where
	\begin{equation}
		\label{Dual}
			\begin{aligned}
			g_{\bm{x},\bm{y}}\left(\bm{\eta}\right) \ &= \ \sup_{\bm{x},\bm{y}} \ \widetilde{L}_{\bm{\eta}}\left(\bm{x},\bm{y}\right)\\
			{\rm s.t.} \ & \ \text{(\ref{P1}a)},\ \text{(\ref{P1}f)}.
			\end{aligned}
	\end{equation}
	Notably, since (\ref{P1.1}) is convex and (\ref{P1.1}a) contains only linear and affine inequalities, according to the duality property~\cite{boyd2004convex}, the above primal-dual transformation w.r.t. $\mathbf{D1.1}$ determines at least the best upper bound of $\mathbf{P1.1}$.
	Hence, our focus now shifts to seeking $\bm{x}^{*}$ and $\bm{y}^{*}$ through solving problem \eqref{Dual} in an iterative fashion of updating $\bm{\eta}$ with a subgradient method~\cite{boyd2003subgradient}.
	
	Before that, all cross terms of $\bm{x}$ and $\bm{y}$ in $\widetilde{L}_{\bm{\eta}}\left(\bm{x},\bm{y}\right)$ need to be tackled for tractability, where $x_{ij}y_{ij}$ and $x_{ij}(1-y_{ij})$ are the only two ways of crossing.
	Combined with constraints (\ref{P1}a) and (\ref{P1}f), here we create a new BS-related index $j' \in \mathcal{J}'=\{1,2,\cdots,J,J+1,J+2,\cdots,2J\}$ and define a new variable set $\bm{\nu}=\{\nu_{ij'}\in \left\{ 0,1\right\}\mid i \in \mathcal{U}, j\in \mathcal{J}'\}$, such that
	\begin{equation}
		\label{newvariableset}
			\nu_{ij'}=\begin{cases}
			x_{ij'}y_{ij'}, & \text{if } j' \in \mathcal{J}=\{1,2,\cdots,J\};\\
			x_{i(j'-J)}(1-y_{i(j'-J)}),& \text{if } j' \in \mathcal{J}'\backslash \mathcal{J}.
		\end{cases}
	\end{equation}
	Among them, $\nu_{ij'}=1$ at $j' \in \mathcal{J}$ represents that MU $i$ selects the SemCom mode to be associated with BS $j'$, and $\nu_{ij'}=1$ at $j' \in \mathcal{J}'\backslash \mathcal{J}$ means that MU $i$ selects the BitCom mode to be associated with BS $(j'-J)$.
	If $\nu_{ij'}=0$, it indicates that MU $i$ is not associated with BS $j'$ (if $j'\in \mathcal{J}$) or BS $(j'-J)$ (if $j'\in \mathcal{J}'\backslash \mathcal{J}$).
	Similarly, we also define a new constant set $\bm{\xi}=\{\xi_{ij'}\mid i \in \mathcal{U}, j\in \mathcal{J}'\}$ to characterize all coefficients of $x_{ij}y_{ij}$ and of $x_{ij}(1-y_{ij})$ in $\widetilde{L}_{\bm{\eta}}\left(\bm{x},\bm{y}\right)$, such that
	\begin{equation}
		\label{newconstantset}
			\xi_{ij'}=\begin{cases}
			\overline{M}_{ij'}^{S_{th}}\!-\!\eta_{j'}z_{ij'}^{S_{th}},& \text{if } j' \in \mathcal{J};\\
			\overline{M}_{i(j'-J)}^{B_{th}}\!-\!\eta_{(j'-J)}z_{i(j'-J)}^{B_{th}},& \text{if } j' \in \mathcal{J}'\backslash \mathcal{J}.
		\end{cases}
	\end{equation}
	
	As such, given the initial dual variable $\bm{\eta}$, problem \eqref{Dual} should be straightforwardly converted to
	\begin{align}
		\mathbf{P1.2}:\ \max_{\bm{\nu}} \quad & \sum_{i\in \mathcal{U}}\sum_{j'\in \mathcal{J}'}\xi_{ij'}\nu_{ij'}~\label{P1.2}\\
		{\rm s.t.} \quad & \sum_{j'\in \mathcal{J}'}\nu_{ij'}= 1,\ \forall i\in \mathcal{U},\tag{\ref{P1.2}a}\\
		& \nu_{ij'}\in \left\{ 0,1\right\},\ \forall \left( i,j'\right) \in\mathcal{U}\times \mathcal{J}'.\tag{\ref{P1.2}b}
	\end{align}
	It is easily derived from $\mathbf{P1.2}$ that for any $i\in \mathcal{U}$, the optimal $j'$ such that $\nu_{ij'}=1$ is exactly the $j'$ that enables the maximum $\xi_{ij'}$ compared with any other $j'\in \mathcal{J}'$.
	In other words, let $\widehat{j'}=\arg \max_{j'\in \mathcal{J}'}\xi_{ij'}, \forall i \in \mathcal{U}$, we can determine $\bm{x}^{*}$ and $\bm{y}^{*}$ for each MU $i$ and BS $j$ in the HSB-Net by
	\begin{equation}
		\label{eachiteration}
		\begin{cases}
			x_{ij}^{*}=1,\ y_{ij}^{*}=1,& \text{if } \widehat{j'} \in \mathcal{J} \text{ and } j=\widehat{j'};\\
			x_{ij}^{*}=1,\ y_{ij}^{*}=0,& \text{if } \widehat{j'}\!\in\! \mathcal{J}'\backslash \mathcal{J} \text{ and } j\!=\!\widehat{j'}\!-\!J;\\
			x_{ij}^{*}\!=\!0,& \text{otherwise}.
		\end{cases}
	\end{equation}
	
	Afterward, the partial derivatives w.r.t. $\bm{\eta}$ in the objective function $H\left(\bm{\eta}\right)$ in $\mathbf{D1.1}$ are set as the subgradient direction in each update iteration.
	Now suppose that in a certain iteration, e.g., iteration $l$, in line with constraint (\ref{D}a), each $\eta_{j}$ ($j \in \mathcal{J}$) is updated as the following rule:
	\begin{equation}
		\label{multiplierupdate}
		\eta_{j}\left(l+1\right)=\max\left\{\eta_{j}\left(l\right)-\epsilon(l)\cdot\nabla H\left(\eta_{j}\right),0\right\},
	\end{equation}
	where
	\begin{equation}
		\label{direction}
		\nabla H\left(\eta_{j}\right)=Z_{j}-\sum_{i\in \mathcal{U}}x_{ij}\left[y_{ij}z_{ij}^{S_{th}}+\left(1-y_{ij}\right)z_{ij}^{B_{th}}\right],
	\end{equation}
	and $\epsilon\left(l\right)$ denotes the stepsize of the update in iteration $l$.
	In general, the convergence of the subgradient descent method can be guaranteed with a properly preset stepsize~\cite{jiang2016optimal}.
	
	Nevertheless, it is worth noting that the above solutions cannot always directly reach the optimality of $\mathbf{P1.1}$, as the BA constraint (\ref{P1.1}a) may be violated at some BSs within each iteration.
	Such violations can affect the convergence of the subgradient method, and may take the obtained solutions out of its feasible region~\cite{boyd2004convex,boostanimehr2014unified}.
	Inspired by~\cite{bertsekas2015parallel}, here we adopt a preference list-based heuristic algorithm to project the solution obtained in each iteration back to the feasible set of (\ref{P1.1}a).
	To be concrete, $\bm{x}^{*}$ and $\bm{y}^{*}$ obtained by \eqref{newvariableset}-\eqref{direction} are first leveraged to identify the index list of the BSs that violate (\ref{P1.1}a), denoted as $\widetilde{\mathcal{J}}=\left\{j\mid j\in\mathcal{J},\sum_{i\in \mathcal{U}}\!x_{ij}^{*}\!\left[y_{ij}^{*}z_{ij}^{S_{th}}\!+\!\left(1\!-\!y_{ij}^{*}\right)z_{ij}^{B_{th}}\!\right]\!> \!Z_{j}\right\}$.
	Consider an arbitrary BS $\tilde{j}\in \widetilde{\mathcal{J}}$, let $\mathcal{U}_{\tilde{j}}\!=\!\{i\mid i\in \mathcal{U},x_{i\tilde{j}}^{*}\!=\!1\}$ store all its current associated MUs, the MU that consumes the largest amount of bandwidth among all MUs can be found by
	\begin{equation}
	\label{mostbandwidthMU}
		\hat{i}=\arg\max_{i\in\mathcal{U}_{\tilde{j}}}\left[y_{i\tilde{j}}^{*}z_{i\tilde{j}}^{S_{th}}+\left(1-y_{i\tilde{j}}^{*}\right)z_{i\tilde{j}}^{B_{th}}\right].
	\end{equation}
	Next, we assume that MU $\hat{i}$ has an initial variable set $\mathcal{J}'_{\hat{i}}=\mathcal{J}'$, which can be reckoned as its UA and MS preference list pertinent to optimizing $\mathbf{P1.2}$.
	Since the solution $(x_{\hat{i}\tilde{j}}^{*}, y_{\hat{i}\tilde{j}}^{*})$ is obviously inapplicable due to the insufficient bandwidth resources at BS $\tilde{j}$, let the corresponding index $j'=\tilde{j}$ in its SemCom case or $j'=\tilde{j}+J$ in its BitCom case be removed from MU $\hat{i}$'s preference list $\mathcal{J}'_{\hat{i}}$.
	That is,
	\begin{equation}
		\mathcal{J}'_{\hat{i}}=\begin{cases}
			\mathcal{J}'_{\hat{i}}\backslash\tilde{j},& \text{if } y_{\hat{i}\tilde{j}}^{*}=1;\\
			\mathcal{J}'_{\hat{i}}\backslash\left(\tilde{j}+J\right),& \text{if } y_{\hat{i}\tilde{j}}^{*}=0,
		\end{cases}
	\end{equation}
	whereby its current optimal $\widehat{j'}$ becomes
	\begin{equation}
	\label{maxjupdate}
		\widehat{j'}=\arg \max_{j'\in \mathcal{J}'_{\hat{i}}}\xi_{\hat{i}j'}.
	\end{equation}
	
	Calculating \eqref{eachiteration} again, it is able to update MU $\hat{i}$'s UA and MS solutions $(x_{\hat{i}j}^{*}, y_{\hat{i}j}^{*})$ over any BS $j\in\mathcal{J}$ as well as BS $\tilde{j}$'s UA list $\mathcal{U}_{\tilde{j}}$.
	After that, the satisfaction of constraint (\ref{P1.1}a) w.r.t. BS $\tilde{j}$ should be rechecked, and even if it is still in violation, we can repeat the operations between \eqref{mostbandwidthMU} to \eqref{maxjupdate} until (\ref{P1.1}a) is eventually met at BS $\tilde{j}$.
	Likewise, the above procedure can be applied to any other BS $j\in \widetilde{\mathcal{J}}$ until the bandwidth constraints are fulfilled at all BSs after each iteration.
	In summary, by alternately updating $(\bm{x},\bm{y})$ and $\bm{\eta}$ in combination with the proposed preference list-based heuristic algorithm, the UA and MS problems can be near-optimally solved in the HSB-Net.
	

	\subsection{Optimal Solution for BA with Complexity Analysis}
	According to the obtained UA solution $\bm{x}^{*}$ and MS solution $\bm{y}^{*}$, we aim to reallocate all bandwidth resources of each BS $j$ ($\forall j \in \mathcal{J}$) to all its associated MUs, thus a total of $S$ BA subproblems w.r.t. $\mathbf{P1}$ are constructed.
	Based on Proposition \ref{Prop2} alongside the preset bandwidth threshold $z_{ij}^{S_{th}}$ and $z_{ij}^{B_{th}}$, each BA subproblem of BS $j$ is formulated as follows:
	\begin{align}
		\mathbf{P1.3}_{j}:\ \max_{\bm{z}} \quad & \sum_{i\in \mathcal{U}_{j}^{S}}\overline{M}_{ij}^{S}+\sum_{i\in \mathcal{U}_{j}^{B}}\overline{M}_{ij}^{B}~\label{P1.3}\\
		{\rm s.t.} \quad & \sum_{i\in \mathcal{U}_{j}^{S}\cup\mathcal{U}_{j}^{B}}z_{ij}=Z_{j},\tag{\ref{P1.3}a}\\
		& z_{ij}\geqslant z_{ij}^{S_{th}},\ \forall i \in\mathcal{U}_{j}^{S},\tag{\ref{P1.3}b}\\
		& z_{ij}\geqslant z_{ij}^{B_{th}},\ \forall i \in\mathcal{U}_{j}^{B},\tag{\ref{P1.3}c}
	\end{align}
	where $\mathcal{U}_{j}^{S}\!=\!\left\{i\mid i\in\mathcal{U},x_{ij}^{*}=1,y_{ij}^{*}=1\right\}$ stands for the set of all SemCom-enabled MUs associated with BS $j$, and $\mathcal{U}_{j}^{B}=\left\{i\mid i\in\mathcal{U},x_{ij}^{*}=1,y_{ij}^{*}=0\right\}$ represents the set of all BitCom-enabled MUs associated with BS $j$.
	Then given the convex property of $\Re_{ij}(\cdot)$~\cite{10261329}, we have the objective function and all constraints of each $\mathbf{P1.3}_{j}$ are convex, thereby efficient linear programming toolboxes such as CVXPY~\cite{diamond2016cvxpy} can be directly applied to obtain the optimal BA solution for the HSB-Net.
	Note that such obtained BA solution will not be used to further optimize $\bm{x}^{*}$ and $\bm{y}^{*}$ due to the bandwidth resource exhaustion.
	
	\subsection{Algorithm Analysis}
	To better demonstrate the full picture of the proposed allocation in HSB-Net, we summarize the relevant technical points and enclose them in the following Algorithm~\ref{Algo1}.
	\begin{breakablealgorithm}
				\caption{The Proposed Resource Allocation in HSB-Net}
				\label{Algo1}
				\begin{algorithmic}[1]
					\REQUIRE \textit{Network parameters $U$, $J$, $T$, $Z_{j}$, $\tau_{i}$, $\rho_{ij}$, $\lambda_{i}$, $L$, $\mu_{i}^{\mathit{Mat}}$, $\mu_{i}^{\mathit{Mis}}$, $F$, $\delta_{0}$, $\theta_{0}$, $\mathit{M}_{i}^{o}$, $\Re_{ij}(\cdot)$ based on pre-trained SemCom models, and $\gamma_{ij}$ with a known probability distribution}
					\ENSURE \textit{UA solution $\bm{x}^*$, MS solution $\bm{y}^*$, and BA solution $\bm{z}^*$}
					\FOR{$i \leftarrow1$ \textit{to} $U$}
						\FOR{$j \leftarrow1$ \textit{to} $J$}
							\STATE \textit{Find $z_{ij}^{S_{th}}$ and $z_{ij}^{B_{th}}$ based on the boundary conditions
							\STATE \quad of constraints (\ref{P1}c)-(\ref{P1}e), as in the context of~\eqref{gestsa}
							\STATE Calculate $\overline{M}_{ij}^{S_{th}}$ and $\overline{M}_{ij}^{B_{th}}$ given the context of $\mathbf{P1.1}$}
						\ENDFOR
					\ENDFOR
					\STATE \textit{Initialize the subgradient iteration index as $l\leftarrow 1$ and all
					\STATE \quad dual variables $\bm{\eta}(1)$ to proper positive values}
					\STATE \textit{Set $V$ as the maximum number of subgradient iteration}
					\WHILE{$l\leqslant V$}
						\FOR{$i \leftarrow1$ \textit{to} $U$}
							\FOR{$j \leftarrow1$ \textit{to} $J$}
								\STATE \textit{Generate $\bm{\xi}(l)$ by~\eqref{newconstantset} for $\mathbf{P1.2}$
								\STATE Determine $x_{ij}^*(l)$ and $y_{ij}^{*}(l)$ by~\eqref{eachiteration}}
							\ENDFOR
						\ENDFOR
						\FOR{$j \leftarrow1$ \textit{to} $J$}
							\IF{\textit{constraint (\ref{P1.1}a) is violated at BS $j$}}
								\STATE \textit{Update $\bm{x}^*(l)$ and $\bm{y}^*(l)$ by~\eqref{eachiteration} and~\eqref{mostbandwidthMU}-\eqref{maxjupdate} until
								\STATE \quad constraint (\ref{P1.1}a) is satisfied at BS $j$} 
							\ENDIF
						\ENDFOR
						\STATE \textit{Update $\bm{\eta}(l+1)$ by~\eqref{multiplierupdate} and~\eqref{direction}}
						\STATE $l\leftarrow l+1$
					\ENDWHILE
					\RETURN $\left(\bm{x}^*,\bm{y}^*\right)\leftarrow\left(\bm{x}^*(V),\bm{y}^*(V)\right)$
					\FOR{$j \leftarrow1$ \textit{to} $J$}
						\STATE \textit{Find $z_{ij}^*$ by CVXPY for all MU $i$ associated with BS $j$}
					\ENDFOR
					\RETURN $\bm{z}^*$
			\end{algorithmic}
		\end{breakablealgorithm}
	
	In terms of the computational complexity of Algorithm~\ref{Algo1}, determining the minimum $z_{ij}$ allocated to each potential UA link first requires $\mathcal{O}(F^2)$ complexity to compute the one-step state transition probability matrix of its PTQ as given in the proof of Proposition~\ref{Prop1}, hence $\mathcal{O}(UJF^2)$ complexity is needed for obtaining $\mathbf{P1.1}$.
	Then, in each iteration of solving $\mathbf{D1.1}$, the complexity is $\mathcal{O}(UJ^2)$ for at most $J$ violated BSs to find their respective largest bandwidth-consumed MUs in a group of $UJ$ variables.
	As such, if let $V$ denote the required number of iterations that leads to convergence of $\mathbf{D1.1}$, finalizing the UA and MS solutions needs a total of $\mathcal{O}(VUJ^2)$ complexity.
	Finally, since each $\mathbf{P1.3}_{j}$ can be solved by the linear programming method with complexity $\mathcal{O}(U^2)$~\cite{lee2019solving}, the proposed wireless resource management solution has a polynomial-time overall complexity of $\mathcal{O}(UJ(F^2+VJ+U))$.
		
	\section{Numerical Results and Discussions}
	\begin{table}[t]
		\centering
		\caption{Simulation Parameters}
		\label{SimuPara}
		\setlength{\tabcolsep}{3pt}
		\renewcommand\arraystretch{1.5}
		\begin{tabular}{|m{4.3cm}<{\raggedright}|m{3.8cm}<{\raggedright}|}\hline
			\textbf{Parameters} & \textbf{Values} \\ \hline
			Bandwidth budget of each BS ($Z_{j}$) & $15$ MHz~\cite{xu2017power}\\ \hline
			Transmit power of each MU & $20$ dBm \\ \hline
			Noise power & $-111.45$ dBm~\cite{boostanimehr2014unified} \\ \hline
			Path loss model & $34+40\log\left(d\ \text{[m]}\right)$ \\ \hline
			Time slot length ($T$) & $1$ ms \\ \hline
			Number of bits in each packet ($L$) & $800$ bits \\ \hline
			Packet buffer size of PTQ ($F$) & $20$ \\ \hline
			Maximum average packet queuing latency threshold ($\delta_{0}$) & $20$ ms \\ \hline
			Maximum average packet loss ratio threshold ($\theta_{0}$) & $0.01$ \\ \hline
		\end{tabular}
	\end{table}
	In this section, numerical evaluations are conducted to demonstrate the performance of our proposed wireless resource management solution in the HSB-Net, where we employ Python 3.7-based PyCharm as the simulator platform and implement it in a workstation PC featuring the AMD Ryzen-9-7900X processor with 12 CPU cores and 128 GB RAM.
	To preserve generality, we first model a circular area with a radius of $300$ meters, in which $200$ MUs and $10$ BSs are randomly dropped.
	Moreover, the SINR $\gamma_{ij}$ follows a Gaussian distribution with standard deviation of $4$ dB~\cite{moustakas2013sinr}.
	For brevity, some simulation parameters not mentioned in the context and their fixed values are summarized in Table~\ref{SimuPara}.
	
	In SemCom-relevant settings, we simulate a general text transmission scenario to examine the proposed solution.
	Such performance test can also be accomplished with other content types like images or videos, and the reason we choose text is to leverage existing natural language processing models that have been well validated in SemCom-related works.
	Particularly, the Transformer in~\cite{xie2021deep} is adopted as the unified semantic encoder for all SemCom links, and the PyTorch-based Adam optimizer is applied for model training with an initial learning rate of $0.001$.
	Based on the public dataset extracted from the proceedings of European Parliament~\cite{koehn2005europarl}, the expression of B2M function $\Re_{ij}(\cdot)$ at each SemCom link can be well approximated from extensive model tests~\cite{10261329}.
	Note that since the B2M did not specify any particular DL model to perform SemCom, other DL models can also be adopted to fit the B2M function without changing other settings and solutions.
	
	As for the queuing modeling part, each MU's average knowledge-matching degree $\tau_{i}$, minimum message rate threshold $\mathit{M}_{i}^{o}$, and BitCom-based B2M coefficient $\rho_{ij}$ are randomly generated in the range of $0.6 \sim 1$, $50 \sim 100$, and $2\times 10^{-5}\sim 2\times 10^{-4}$, respectively.
	Besides, the average packet arrival rate $\lambda_{i}$ is prescribed at $1000$ packets/s for all MUs~\cite{7434039}, while the average interpretation times of knowledge-matching and -mismatching packets in SCQ are considered as $8\times 10^{-4}$ and $1\times 10^{-3}$ s/packet, respectively.
	Furthermore, we set a dynamic stepsize of $\epsilon(l)= 1\times 10^{-6}/l$ to update the Lagrange multipliers in~(\ref{multiplierupdate}), where the convergence of each trial can be always guaranteed.
	It is worth mentioning that all the above parameter values are set by default unless otherwise specified, and all subsequent numerical results are obtained by averaging over a sufficiently large number of trials.
	
	For comparison purposes, here we employ four different resource management benchmarks in HSB-Nets by combining the max-SINR UA scheme (i.e., each MU is associated with the BS enabling the strongest SINR) with several differing MS and BA schemes, respectively.
	To the best of our knowledge, no existing work has proposed any benchmark solutions dedicatedly for MS, and therefore, two heuristic schemes are developed as MS baselines: (MS-I) A~\textit{knowledge matching degree-based} method, where each MU selects the SemCom mode when its knowledge matching degree is above a preset threshold (e.g., a threshold of $0.8$ has been used in our simulations), and otherwise selects the BitCom mode; (MS-II) A~\textit{SINR-based} method, where each MU selects the BitCom mode when its SINR is above a preset threshold (e.g., a threshold of $6$ dB has been used in our simulations), and otherwise selects the SemCom mode.\footnote{The MS-I scheme is intuitive since the higher the knowledge matching degree, the better the semantic-related performance~\cite{10261329}. As for the MS-II scheme, this is because SemCom shows more powerful anti-noise capability in the low-SINR region~\cite{xie2021deep}, while BitCom ensures higher content transmission accuracy in the high-SINR region~\cite{xia2023wiservr}, thereby MS-II should be applicable.}
	In parallel, two typical BA schemes are adopted as baselines: (BA-I) The~\textit{water-filling} algorithm~\cite{he2013water}; (BA-II) The~\textit{evenly-distributed} algorithm~\cite{ye2013user}.
	
	\begin{figure}[t]
		\centering
		\includegraphics[width=0.4\textwidth]{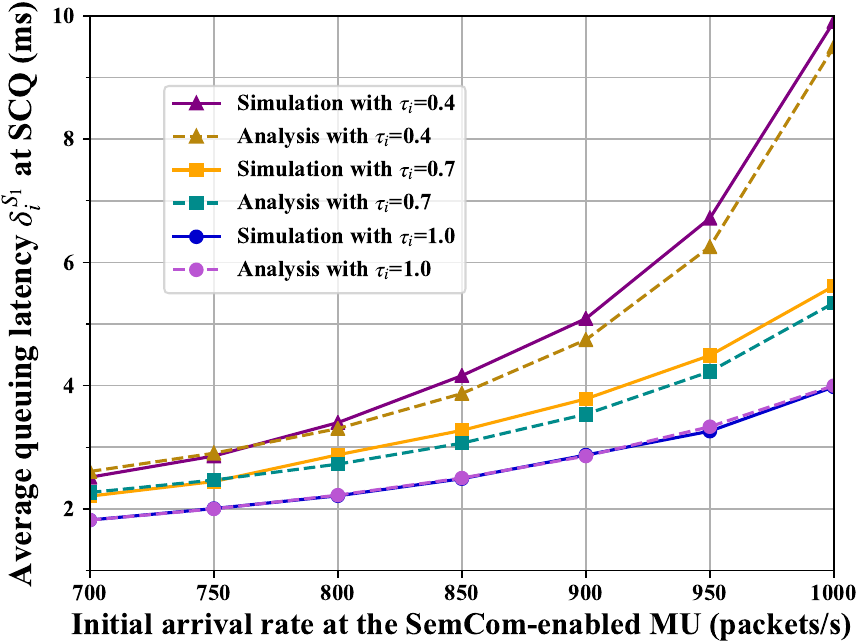} 
		\caption{Simulated and analytical results w.r.t. average queuing latency $\delta_{i}^{S_1}$ at SCQ for varying packet arrival rates and average knowledge-matching degrees.}
		\label{SimuFigure1}
    \end{figure}
    
    \subsection{Queuing Model Validation}
    For starters, we simulate the entire packet queuing processes for SCQ and PTQ at a SemCom-enabled MU with a default average knowledge-matching degree $\tau_{i}=1.0$ and a default SINR $\gamma_{ij}=0$ dB to validate the analytical accuracy of the derived queuing model.
    In detail, the analysis results are based on the direct computation of average packet queuing latency and packet loss ratio as in~(\ref{SCQ_queuing_delay})-(\ref{PTQ_queuing_delay}).
    In contrast, the simulation results are calculated by generating various randomized processes (including Poisson packet arrival, general-distribution based SCQ-packet departure and SINR-stochasticity based PTQ-packet departure) and averaging over $10,000$ trials.
    
    Figure~\ref{SimuFigure1} first depicts the average queuing latency $\delta_{i}^{S_1}$ at SCQ by increasing the initial packet arrival intensity $\lambda_i$ from $750$ to $1050$ packets/s, where $\tau_{i}=0.4$, $0.7$, and $1.0$ are all taking into account.
    It is seen that the analytical curve basically agrees with the simulated one as $\lambda_i$ grows, and the higher the $\tau_{i}$, the closer the two latency curves in values.
    This can be explained by that the lower $\tau_{i}$ indicates the worse semantic inference capability for packet departure at SCQ, resulting in more uncertainty, i.e., higher fluctuation, on each randomly generated semantic-coding time.
    However, in our queuing analysis, the semantic-coding times of all knowledge-mismatching packets are simply approximated to have the same rate of $1/\mu_{i}^{\mathit{Mis}}$, which ignores the discrepancy between different knowledge-matching degrees, and thus rendering the numerical bias between simulated and analytical results in the lower $\tau_{i}$ region.
    Besides, the average queuing latency increases with the packet arrival rate in each case, which trend is obvious as the semantic-coding efficiency is fixed at SCQ. 
    
     \begin{figure}[t]
		\centering
		\includegraphics[width=0.4\textwidth]{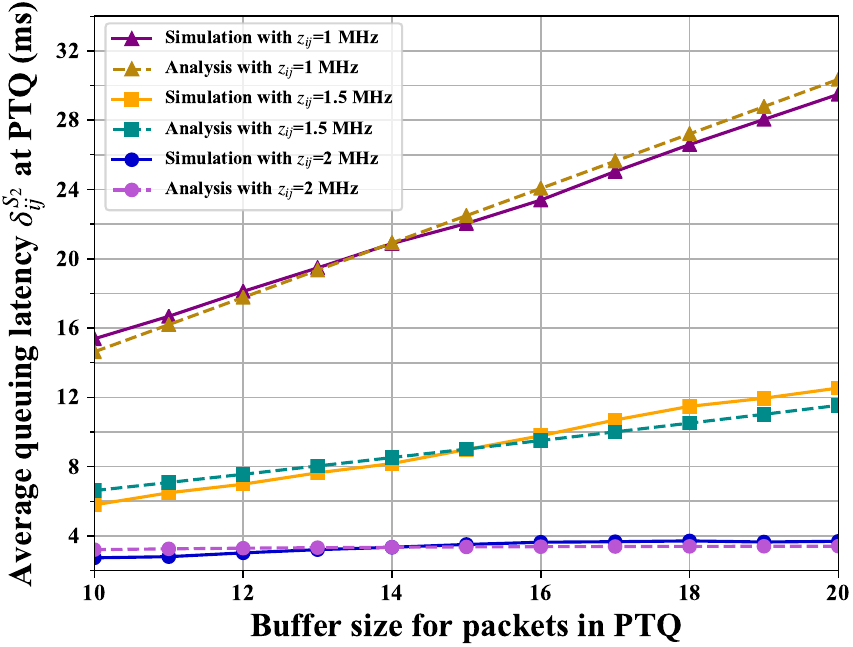} 
		\caption{Simulated and analytical results w.r.t. average queuing latency $\delta_{ij}^{S_2}$ at PTQ for varying packet buffer sizes and allocated bandwidth resources.}
		\label{SimuFigure2}
    \end{figure}
    
    \begin{figure}[t]
		\centering
		\includegraphics[width=0.4\textwidth]{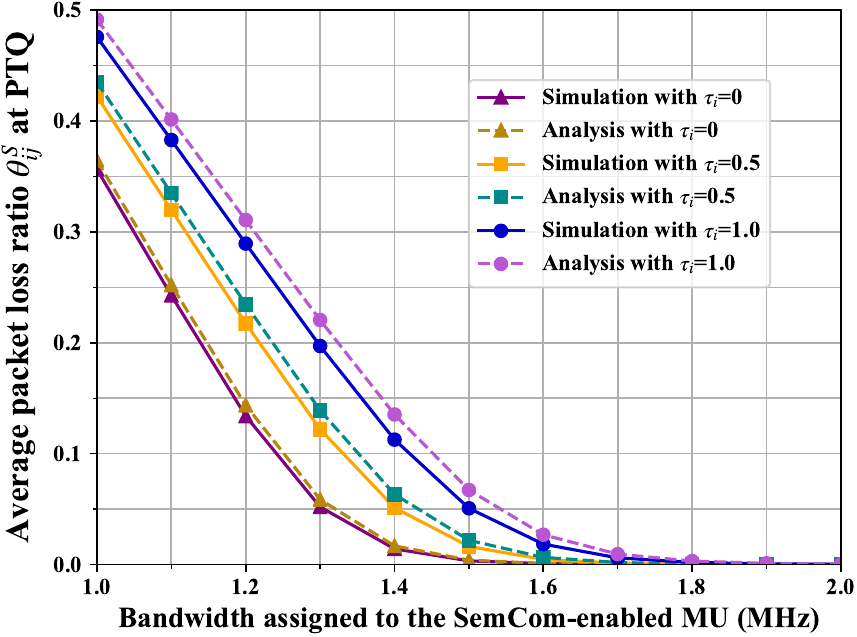} 
		\caption{Simulated and analytical results w.r.t. average packet loss ratio $\theta_{ij}^{S}$ at PTQ for varying bandwidth resources and knowledge-matching degrees.}
		\label{SimuFigure3}
    \end{figure}
    Next, we compare the simulated and analytical results of PTQ in terms of its average queuing latency and packet loss ratio in Fig.~\ref{SimuFigure2} and Fig.~\ref{SimuFigure3}, respectively, which are basically consistent in both cases.
    By varying PTQ's buffer size $F$ from $10$ to $22$, Fig.~\ref{SimuFigure2} shows a moderate increasing trend in average queuing latency $\delta_{ij}^{S_2}$ with different allocated bandwidth $z_{ij}=1$, $1.5$, and $2$ MHz.
    This is logical since the buffer with a larger size is more likely to hold a long queue length, resulting in more average waiting time per packet according to~(\ref{PTQ_queuing_delay}).
    Moreover, it can be observed that the less the bandwidth assigned to the MU, the higher the $\delta_{ij}^{S_2}$ while the steeper the upward trend.
    Herein, the former phenomenon is reasonable due to the low packet departure rate as in~(\ref{PTQDepart}).
    The latter is because that as the given $z_{ij}$ grows, the rapid departure of packets gradually dominates the queuing process of PTQ, thereby the small changes in the buffer size could only have a slight impact on the rendered $\delta_{ij}^{S_2}$ performance.
   	
   	Meanwhile, Fig.~\ref{SimuFigure3} presents the average packet loss ratio $\theta_{ij}^{S}$ at PTQ versus the allocated bandwidth $z_{ij}$ under three average knowledge-matching degrees of $\tau_{i}=0$, $0.5$, and $1.0$, where the simulated results can always fit the analytical ones well.
   	Specifically, the obtained $\theta_{ij}^{S}$ first decreases with $z_{ij}$, and then converges close to $0$ when $z_{ij}$ exceeds around $1.8$ MHz.
    The rationale behind this is similar to Fig.~\ref{SimuFigure2}, i.e., the packets arriving at the PTQ with a higher departure rate are less likely to be blocked.
    Furthermore, combined~(\ref{SPTQmean}) with the setting of $\mu_{i}^{\mathit{Mat}}\!>\!\mu_{i}^{\mathit{Mis}}$, it is seen that the higher the $\tau_{i}$, the higher the packet arrival rate of PTQ, and thus the greater the likelihood that its buffer tends to be full.
    Notably, the average packet loss ratio of PTQ and the overall queuing latency of both SCQ and PTQ should be taken into account together to meet the preset delay and reliability requirements.
    For instance, $\theta_{ij}^{S}$ can reach the threshold of $\theta_{0}=0.01$ by assigning $1.55$ MHz bandwidth to the same MU with $\tau_{i}=0.5$.
    However, even the default $\lambda_i = 1000$ packets/s will cause the queuing delay of $9.1$ ms at SCQ and $11.5$ ms at PTQ (i.e., the total of $20.6$ ms) in the same case, which has exceeded the threshold $\delta_{0}=20$ ms.
   
   \subsection{Performance of the Proposed Solution}
	\begin{figure}[t]
		\centering
		\includegraphics[width=0.4\textwidth]{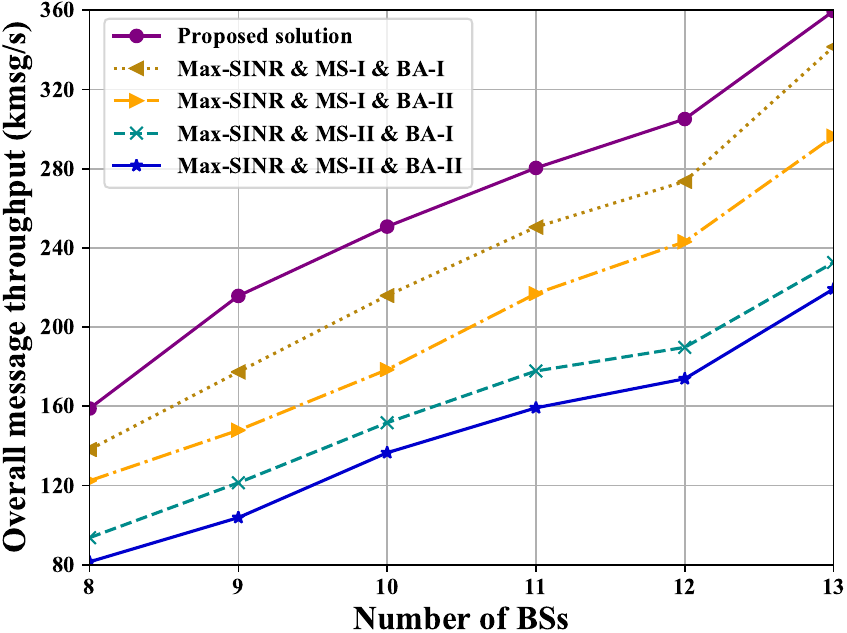} 
		\caption{Time-averaged overall message throughput (\textit{kmsg/s}) versus different numbers of BSs in the HSB-Net.}
		\label{SimuFigure4}
    \end{figure}
    
\begin{figure}[t]
		\centering
		\includegraphics[width=0.4\textwidth]{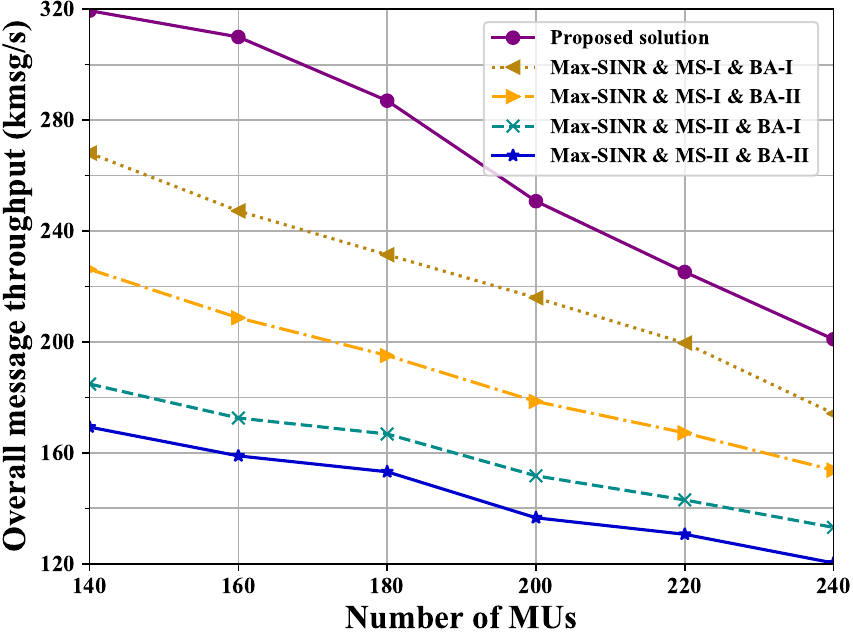} 
		\caption{Time-averaged overall message throughput (\textit{kmsg/s}) versus different numbers of MUs in the HSB-Net.}
		\label{SimuFigure5}
    \end{figure}

    To validate our proposed resource management solution, we test the overall message throughput of HSB-Net under different numbers of BSs and MUs in Fig.~\ref{SimuFigure4} and Fig.~\ref{SimuFigure5}, respectively, in comparison with the four benchmarks.
    As first elucidated in Fig.~\ref{SimuFigure4}, by varying the number of BSs from $8$ to $13$, the message throughput performance of the proposed solution gradually increases from around $160$ to $360$ kmsg/s ($1$ kmsg/s $=$ $1000$ msg/s), and consistently outperforms these benchmarks.
    For example, a performance gain of the proposed solution is about $29.9$ kmsg/s compared with the benchmark of Max-SINR plus MS-I plus BA-I and $102.6$ kmsg/s compared with the benchmark of Max-SINR plus MS-II plus BA-I when $11$ BSs are located in the HSB-Net.
    Here, such an uptrend is apparent since more BSs represent that more bandwidth resources are available for MUs to achieve higher message rates.
    Particularly in such an uplink scenario of HSB-Net, the increase in the number of BSs does not have any impact on channel interference, and hence a stable growth is observed.
    
    By contrast, Fig.~\ref{SimuFigure5} demonstrates a downward trend of message throughput performance when rising the number of MUs from $140$ to $240$.
    To be concrete, the overall network performance is already saturated at the very beginning in holding $140$ MUs and then deteriorates with the addition of MUs, as the effect of severe channel interference from excessive MUs starts to dominate the more availability of resources.
	In the meantime, it can be seen that our solution still surpasses all the four benchmarks with a significant performance gain.
    For instance, with $160$ MUs in the HSB-Net, the proposed solution realizes a message throughput of about $310$ kmsg/s, i.e., $1.5$ times that of the Max-SINR plus MS-I plus BA-II scheme and $2$ times that of the Max-SINR plus MS-II plus BA-II scheme.

    \begin{figure}[t]
		\centering
		\includegraphics[width=0.4\textwidth]{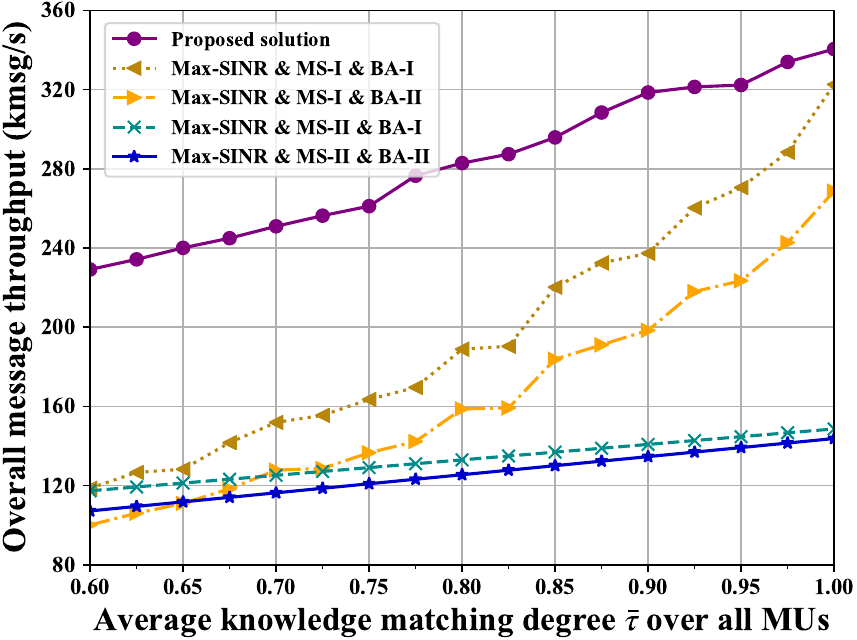} 
		\caption{Time-averaged overall message throughput (\textit{kmsg/s}) versus different average knowledge-matching degrees over all $200$ MUs in the HSB-Net.}
		\label{SimuFigure6}
    \end{figure}
    
    In addition, we compare the message throughput performance with varying overall average knowledge-matching degree $\bar{\tau}=\frac{1}{U}\sum_{i \in \mathcal{U}}\tau_{i}$ as shown in Fig.~\ref{SimuFigure6}.
    Again, our solution still outperforms these benchmarks with the considerable performance gain, especially in the low $\bar{\tau}$ region.
    Besides, a growing message throughput is observed by all solutions as $\bar{\tau}$ increases, and our solution and the MS-I scheme are more affected by changes in $\bar{\tau}$ compared to the MS-II.
    The former trend is intuitive since the larger $\bar{\tau}$ means that there is a greater likelihood for the HSB-Net having MUs with the high B2M transformation rates.
    The latter is first due to the message-throughput-priority design in our objective function~(\ref{P1}), and therefore, our solution is more likely to generate more SemCom-enabled MUs with larger $\bar{\tau}$.
    Likewise, more SemCom-enabled MUs can exist in the same case according to the prescribed MS-I scheme, while the number of SemCom-enabled MUs is only affected by SINR in MS-II, and thus keeps stable irrespective of the change in $\bar{\tau}$.
    
     \begin{figure}[t]
		\centering
		\includegraphics[width=0.4\textwidth]{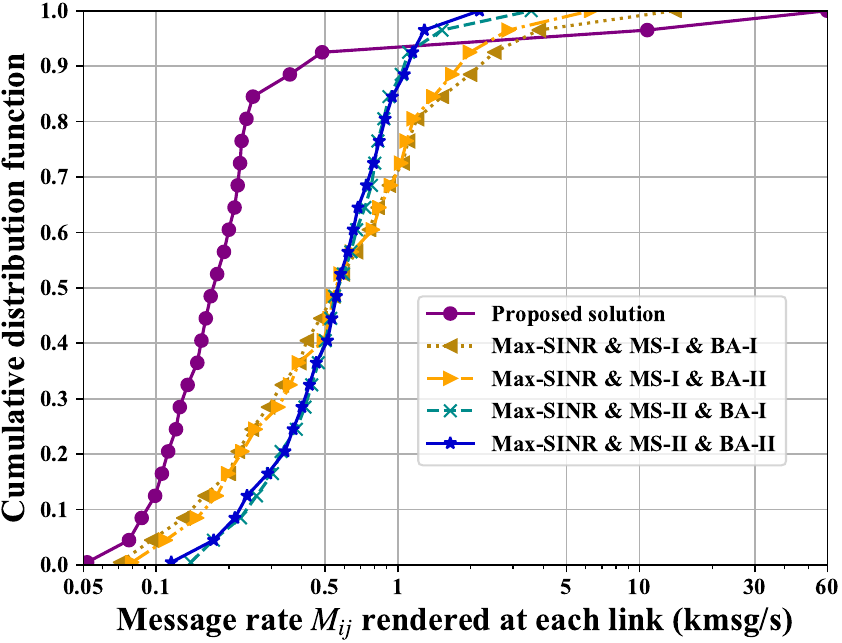} 
		\caption{The CDF of the message rate $M_{ij}$ obtained by the associated link.}
		\label{SimuFigure7}
    \end{figure}
	
	Finally, the CDFs of the message rate $M_{ij}$ rendered at all links are plotted in Fig.~\ref{SimuFigure7}.
	Although most MUs in our solution only get the lower message rates compared with these benchmarks, this is reasonable since our optimization to $\mathbf{P1}$ focuses on the maximization for overall message throughput of all MUs in the HSB-Net.
	Hence, it can be interpreted as that the proposed solution choose to sacrifice user semantic fairness in favor of devoting more bandwidth resources to a smaller number of MUs with better average knowledge-matching degrees, B2M transformation, and SINRs.
	
	\section{Conclusions}
	In this paper, we investigated the wireless resource management problem in a novel yet practical network scenario, i.e., HSB-Net, where SemCom and BitCom modes are available for selection by all MUs.
	To measure SemCom and BitCom with the same performance metric, a B2M transformation function was first introduced to identify the message throughput of each associated link.
	Then, considering the unique semantic coding and knowledge matching mechanisms in SemCom, we modelled a two-stage tandem queuing system for the transmission of semantic packets, followed by the theoretical derivation for average packet loss ratio and queuing latency.
	On this basis, a joint optimization problem was formulated to maximize the overall message throughput of HSB-Net.
	Afterward, a Lagrange primal-dual method was employed and a preference list-based heuristic algorithm was developed to seek the optimal UA, MS, and BA solutions with the low computational complexity.
	Numerical results finally validated the accuracy of our queuing analysis and the performance superiority of our proposed solution in terms of overall message throughput compared with four different benchmarks.
	
	This paper can serve as a pioneer work to offer valuable insights for follow-up research on hybrid SemCom and BitCom networks.
	Other relevant networking issues in the HSB-Net, such as communication mode switching or semantic fairness-driven power or resource block allocation, inevitably arise, which can treat this paper as the fundamental theoretical framework for reference.
	Since this work is limited to the long-term network optimization under known background knowledge conditions of MUs, further problems about instantaneous decision-making for MS and BA with unknown knowledge-matching degrees could be the next research direction.
	
	\begin{appendices}
		\section{Proof of Proposition 1}
		 It is first found from \eqref{PTQevolution} that given any queue length $a$ at slot $t$ (i.e., $Q_{ij}\left(t\right)=a$, $0\leqslant a\leqslant F$), $Q_{ij}\left(t+1\right)$ is determined only by $A'_{i}\left(t\right)$ and $D_{ij}\left(t\right)$.
		Apparently, the stochastic $Q_{ij}\left(t\right)$ across all slots forms a discrete-time Markov process, herein we define $\omega^{a\looparrowright b}_{ij}\left(t\right)=\Pr\left\{Q_{ij}\left(t+1\right)=b\mid Q_{ij}\left(t\right)=a\right\}$ as its one-step state transition probability from state $a$ at slot $t$ to state $b$ at slot $t+1$, $0\leqslant b\leqslant F$.
		Since the PMFs of both $A'_{i}\left(t\right)$ and $D_{ij}\left(t\right)$ are independent of $t$, as mentioned earlier, we re-denote them by $A'_{i}$ and $D_{ij}$ for brevity, respectively.
		As such, $\omega^{a\looparrowright b}_{ij}\left(t\right)$ can be expressed as $\omega^{a\looparrowright b}_{ij}$ as well.
		
		Having these, we have the one-step state transition probability matrix of SemCom-enabled PTQ as
		\begin{equation}
		\label{omegamatrix}
		\bm{\Omega}_{ij}=\left[
			\begin{matrix}
				\omega^{0\looparrowright 0}_{ij} &\omega^{0\looparrowright 1}_{ij}&\cdots & \omega^{0\looparrowright F}_{ij}\\
				\omega^{1\looparrowright 0}_{ij} &\omega^{1\looparrowright 1}_{ij} &\cdots & \omega^{1\looparrowright F}_{ij} \\
				\vdots & \vdots &\ddots & \vdots\\
				\omega^{F\looparrowright 0}_{ij} & \omega^{F\looparrowright 1}_{ij} & \cdots & \omega^{F\looparrowright F}_{ij}
			\end{matrix}
			\right],
		\end{equation}
		where each $\omega^{a\looparrowright b}_{ij}$ can be explicitly calculated in \eqref{state_transition}, placed at the bottom of the next page.
		\begin{figure*}[hb]
		\centering
		\hrulefill
		\begin{equation}
			\omega^{a\looparrowright b}_{ij}\!=\!\begin{cases}
				\!\Pr\left\{A'_{i}=b\right\}, &\text{if }a=0, \ 1\leqslant b\leqslant F-1;\\
				\!\sum_{k=F}^{\infty}\Pr\left\{A'_{i}=k\right\},&\text{if }a=0, \ b=F;\\
				\!\Pr\left\{A'_{i}=0\right\}\sum_{k=a}^{\infty}\Pr\left\{D_{ij}=k\right\},&\text{if }0\leqslant a\leqslant F,\ b=0;\\
				\!\Pr\left\{A'_{i}=b\right\}\sum_{k=a}^{\infty}\Pr\left\{D_{ij}=k\right\}+\sum_{l=0}^{b-1} \Pr\left\{A'_{i}=l\right\}\Pr\left\{D_{ij}=a-b+l\right\},&\text{if }1\leqslant a\leqslant F, 1\leqslant b\leqslant a, b \neq F;\\
				\!\Pr\left\{A'_{i}=b\right\}\sum_{k=a}^{\infty}\Pr\left\{D_{ij}=k\right\}+\sum_{l=0}^{a-1} \Pr\left\{D_{ij}=l\right\}\Pr\left\{A'_{i}=b-a+l\right\},&\text{if }1\leqslant a<b\leqslant F-1;\\
				\!\sum_{k=b}^{\infty}\!\Pr\!\left\{\!A'_{i}\!=\!k\!\right\}\!\sum_{l=a}^{\infty}\!\Pr\!\left\{\!D_{ij}\!=\!l\right\} \!+\!\sum_{l=0}^{a-1}\!\left(\!\Pr\!\left\{\!D_{ij}\!=\!l\right\}\!\sum_{k=b-a+l}^{\infty}\!\Pr\!\left\{\!A'_{i}\!=\!k\!\right\}\!\right),&\text{if }1\leqslant a\leqslant F,\ b=F.\label{state_transition}
			\end{cases}
		\end{equation}
		\end{figure*}
		
		Further noticing that for the queue state transited from $Q_{ij}(t)=0$ to $Q_{ij}(t+1)=0$, the transition probability is
		\begin{equation}
		\begin{aligned}
			\Pr\{Q_{ij}&(t+1)=0\mid Q_{ij}(t)=0\}=\Pr\{A_{i}'=0\}\\
			&=\exp{\left(-\tau_{i}\mu_{i}^{\mathit{Mat}}T-(1-\tau_{i})\mu_{i}^{\mathit{Mis}}T\right)}>0,
			\end{aligned}
		\end{equation}
		which proves that $Q_{ij}(t)=0$ is aperiodic.
		Besides, combined with a fact that each $\omega^{a\looparrowright b}_{ij}$ is time independent and each $Q_{ij}(t)$ has a finite state space, $\{Q_{ij}(t) \mid t=1, 2, \cdots, N\}$ is time-homogeneous, irreducible, and aperiodic.
		Therefore, according to~\cite{bolch2006queueing}, there must be a unique steady-state probability vector $\bm{\alpha}_{ij}=\left[\alpha_{ij}^{0},\alpha_{ij}^{1},\cdots,\alpha_{ij}^{F}\right]^{T}$, which can be obtained by solving
		\begin{equation}
			\label{solveprobvec}
			\bm{\Omega}_{ij}^{T}\bm{\alpha}_{ij}=\bm{\alpha}_{ij} \quad \text{and} \quad \sum_{k=0}^{F}\alpha_{ij}^{k}=1.
		\end{equation}
		This completes the proof.
			
		\section{Proof of Proposition 2}
		It is worth noting in the first place that here we only show the proof in the SemCom-enabled queuing model case of $y_{ij}=1$ for exemplification (given any pair of MU $i$ and BS $j$), since the proof in the BitCom case can be similarly derived based on their analogous modeling for PTQ.
			Notice that $\delta_{ij}^{S_{1}}$ is a known constant as in \eqref{SCQ_queuing_delay}, $\delta_{ij}^{S_{2}}$ in \eqref{PTQ_queuing_delay} and $\theta_{ij}^{S}$ in \eqref{SemComplr} become the only two factors that $z_{ij}$ can influence.
			Further combining that $A_{i}'$ is irrelevant with $z_{ij}$ as in~(\ref{PTQmixedarrival}), let us at first present a lemma of how $z_{ij}$ relates the distribution of $D_{ij}$.
			
			\textit{Lemma 1: The CDF of $D_{ij}$ decreases as $z_{ij}$ increases.}
			
			To prove Lemma 1, we first derive the CDF of $D_{ij}$ from its PMF given in \eqref{PTQDepart} as follows:
			\begin{equation}
				\label{CDFofD}
				\begin{aligned}
					\Pr&\left\{D_{ij}\leqslant k\right\}=\sum_{f=0}^{k}\Pr\left\{D_{ij}= f\right\}\\
					&=\Pr\!\left\{\!\gamma_{ij}\leqslant 2^{\frac{(k+1)L}{Tz_{ij}}}\!-\!1\!\right\}\!-\!\Pr\!\left\{\!\gamma_{ij}\leqslant 2^{\frac{kL}{Tz_{ij}}}\!-\!1\!\right\}\\
					& \qquad +\Pr\!\left\{\!\gamma_{ij}\leqslant 2^{\frac{kL}{Tz_{ij}}}\!-\!1\!\right\}\!-\!\Pr\!\left\{\!\gamma_{ij}\leqslant 2^{\frac{(k-1)L}{Tz_{ij}}}\!-\!1\!\right\}\\
					& \quad \qquad +\Pr\!\left\{\!\gamma_{ij}\leqslant 2^{\frac{(k-1)L}{Tz_{ij}}}\!-\!1\!\right\}\!-\!\Pr\!\left\{\!\gamma_{ij}\leqslant 2^{\frac{(k-2)L}{Tz_{ij}}}\!-\!1\!\right\}\\
					& \qquad \qquad + \cdots \cdots +\Pr\left\{\gamma_{ij}\leqslant 2^{\frac{L}{Tz_{ij}}}\!-\!1\right\}\!-\!\Pr\left\{\gamma_{ij}\leqslant 0\right\}\\
					&=\Pr\!\left\{\!\gamma_{ij}\leqslant 2^{\frac{(k+1)L}{Tz_{ij}}}\!-\!1\!\right\}\!-\!\Pr\!\left\{\gamma_{ij}\leqslant 0\right\}, \ k=0, 1, 2, \cdots,
				\end{aligned}
			\end{equation}
			where slot index $t$ is omitted from all notations associated with the SINR $\gamma_{ij}$ for brevity due to its independence w.r.t. $t$ as aforementioned.
			Given arbitrary known CDF of $\gamma_{ij}$, which is independent with $z_{ij}$, we clearly have that $\Pr\left\{D_{ij}\leqslant k\right\}$ is a monotonically decreasing function of $z_{ij}$.
			This also implies that $\Pr\left\{D_{ij}\geqslant k\right\}$ monotonically increases w.r.t. $z_{ij}$.
			
			Now, let us consider two complementary queuing state subspaces of queue length $Q_{ij}$, denoted by $\overleftarrow{\mathcal{F}_{c}}=\{0, 1, 2, \cdots, c\}$ and $\overrightarrow{\mathcal{F}_{c}}=\{c+1, c+2, \cdots, F\}$, $c=0,1, 2,\cdots,F-1$.
			Given any current state $c$, it can only transit to either a smaller state in $\overleftarrow{\mathcal{F}_{c}}$ or a larger state in $\overrightarrow{\mathcal{F}_{c}}$ in the next step, and the probabilities of the two transition cases occurring sum to $1$.
			According to the one-step transition probability $\omega^{a\looparrowright b}_{ij}$ expressed in \eqref{state_transition}, the probability of state $c$ transiting to any state in $\overleftarrow{\mathcal{F}_{c}}$ should be computed by
			\begin{equation}
				\label{smallerstate}
				\begin{aligned}
					&\ \!\omega^{c\looparrowright 0}_{ij}+\omega^{c\looparrowright 1}_{ij}+\cdots \cdots +\omega^{c\looparrowright c}_{ij}\\
					=\ & \Pr\left\{A_{i}'=0\right\} +\Pr\left\{A_{i}'=1\right\}\sum_{l=1}^{\infty}\Pr\left\{D_{ij}=l\right\}\\
					& \quad  + \Pr\left\{A_{i}'=2\right\}\sum_{l=2}^{\infty}\Pr\left\{D_{ij}=l\right\}\\
					& \quad \qquad + \cdots \cdots + \Pr\left\{A_{i}'=c\right\}\sum_{l=c}^{\infty}\Pr\left\{D_{ij}=l\right\}\\
					=\ & \Pr\left\{A_{i}'=0\right\}+\sum_{k=1}^{c}\left(\Pr\left\{A_{i}'=k\right\}\sum_{l=k}^{\infty}\Pr\left\{D_{ij}=l\right\}\right)\\
					=\ & \Pr\left\{A_{i}'=0\right\}+\sum_{k=1}^{c}\left(\Pr\left\{A_{i}'=k\right\}\Pr\left\{D_{ij}\geqslant k\right\}\right).
				\end{aligned}
			\end{equation}
			According to Lemma 1, (\ref{smallerstate}) is clearly a monotonically increasing function of $z_{ij}$ due to its $\Pr\left\{D_{ij}\geqslant k\right\}$ term.
			In other words, we have that the probability of any fixed state $c$ transiting to a state in $\overrightarrow{\mathcal{F}_{c}}$ monotonically decreases w.r.t. $z_{ij}$.
			Further combined with the obtained steady-state probability vector $\bm{\alpha}_{ij}$, if denoting the cumulative distribution of the queuing system staying in the state space $\overleftarrow{\mathcal{F}_{c}}$ as $W_{ij}^{(c)}=\sum_{l=0}^{c}\alpha_{ij}^{l}$, $W_{ij}^{(c)}$ is increasing w.r.t. $z_{ij}$ for any $c$ as well.
			
			 By leveraging a fact that $W_{ij}^{(F)}=1$, let us first rephrase the numerator term of $\delta_{ij}^{S_{2}}$ in \eqref{PTQ_queuing_delay} as follows:
			 \begin{equation}
			 	\begin{aligned}
			 		\mathds{E}&\left[Q_{ij}(t)\right]=\sum_{k=1}^{F}\alpha_{ij}^{k}+\sum_{k=2}^{F}\alpha_{ij}^{k}+\cdots+\sum_{k=F-1}^{F}\alpha_{ij}^{k}+\alpha_{ij}^{F}\\
			 		&\quad =\left(1-W_{ij}^{(0)}\right)+\left(1-W_{ij}^{(1)}\right)+ \cdots +\left(1-W_{ij}^{(F-1)}\right),
			 	\end{aligned}
			 \end{equation}
			whereby the conclusion that $\mathds{E}\left[Q_{ij}(t)\right]$ is monotonically decreasing w.r.t. $z_{ij}$ holds.
			
			Regarding $\theta_{ij}^{S}$ in \eqref{Drop}, which is also served as the key term in the denominator of $\delta_{ij}^{S_{2}}$, we restructure the formula by highlighting all its implicit terms that transform to $\Pr\left\{D_{ij}\leqslant k\right\}$ and $W_{ij}^{(c)}$, and obtain
			\begin{equation}
				\begin{aligned}
					&G_{ij}\!=\!\sum_{f=1}^{F}\Pr\!\left\{\!A_{i}'\!=\!f\!\right\}\!\left[\sum_{k=0}^{f-1}\Pr\left\{D_{ij}\!\leqslant\! k\right\}\!\left(1\!-\!W_{ij}^{(F-f+k)}\right)\!\right]\\
					&\quad +\! \!\sum_{f\!=\!F+1}^{\infty}\!\!\!\Pr\!\left\{\!A_{i}'\!=\!f\!\right\}\!\left[\!(f\!-\!F)\!+\! \!\sum_{k=0}^{F-1}\!\Pr\!\left\{D_{ij}\!\leqslant\! k\right\}\!\left(\!1\!-\!W_{ij}^{(k)}\!\right)\!\right].
				\end{aligned}
			\end{equation}
			Again employing Lemma 1, we have that $G_{ij}$ monotonically decreases w.r.t. $z_{ij}$, thereby $\theta_{ij}^{S}$ and $\delta_{ij}^{S_{2}}$ should have the same decreasing property.
			Finally, note that $\delta_{ij}^{S}=\delta_{ij}^{S_{1}}+\delta_{ij}^{S_{2}}\geqslant \delta_{ij}^{S_{1}}>0$ always holds in practice, $\delta_{ij}^{S}$ must be monotonically non-increasing w.r.t. $z_{ij}$, which completes the proof.

	\end{appendices}

	
	\bibliographystyle{IEEEtran}
	\bibliography{main}
	
	\begin{IEEEbiography}[{\includegraphics[width=1in, height=1.25in, clip, keepaspectratio]{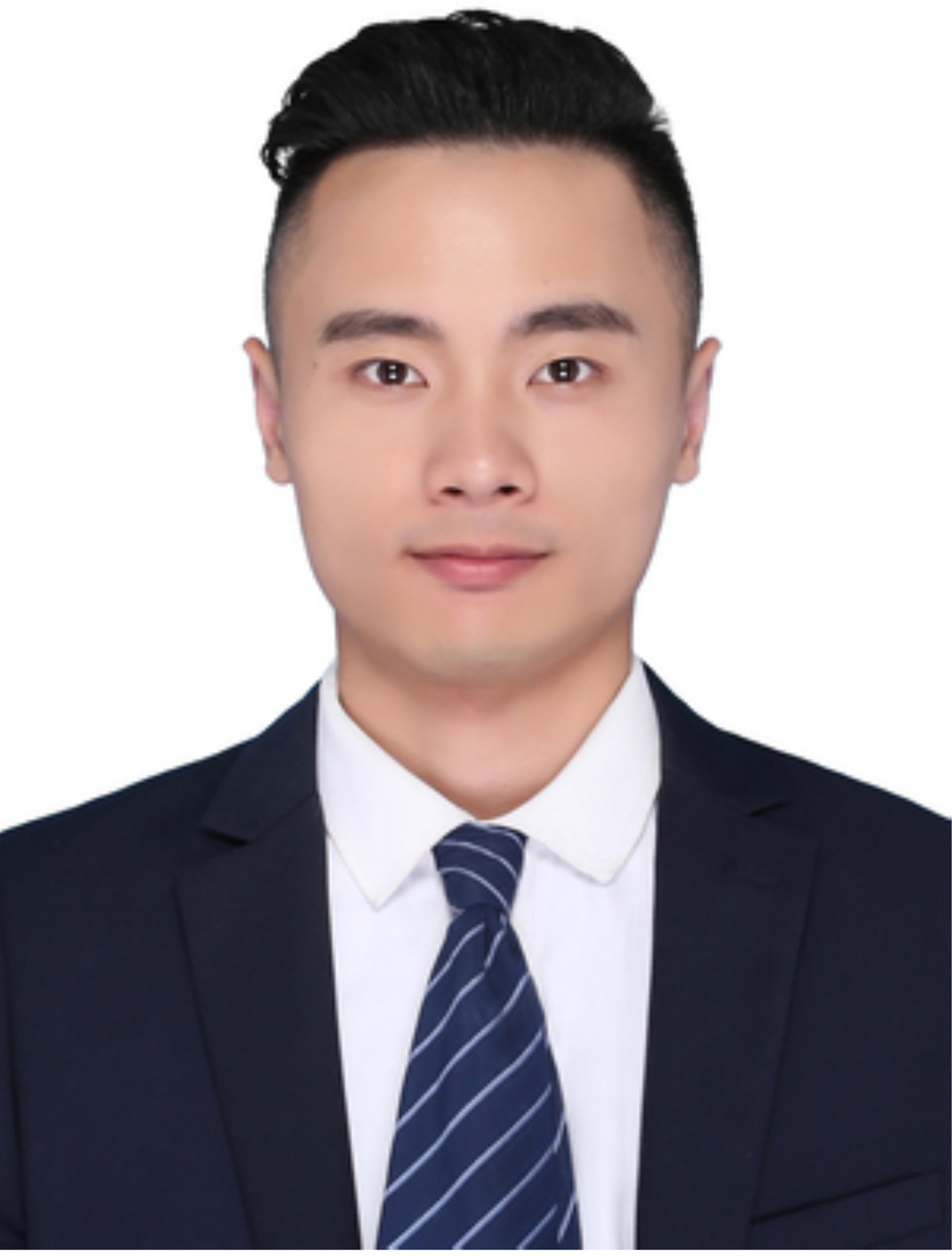}}]{Le Xia} (Member, IEEE)
	 obtained his Ph.D. degree in Electronics and Electrical Engineering from the University of Glasgow, United Kingdom, in 2024. Before that, he received his B.Eng. and M.Eng. in Electronics and Communication Engineering from the University of Electronic Science and Technology of China in 2017 and 2020, respectively. His research interests include next-generation wireless networking, semantic communications, resource optimization, and smart vehicular networks.
	\end{IEEEbiography}

	\begin{IEEEbiography}[{\includegraphics[width=1in, height=1.25in, clip, keepaspectratio]{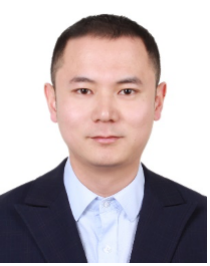}}]{Yao Sun} (Senior Member, IEEE)
	is currently a Lecturer with James Watt School of Engineering, the University of Glasgow, Glasgow, UK. Dr Sun has won the IEEE Communication Society of TAOS Best Paper Award in 2019 ICC, IEEE IoT Journal Best Paper Award 2022 and Best Paper Award in 22nd ICCT. His research interests include intelligent wireless networking, semantic communications, blockchain system, and resource management in next generation mobile networks. Dr. Sun is a senior member of IEEE.
	\end{IEEEbiography}

	\begin{IEEEbiography}[{\includegraphics[width=1in, height=1.25in, clip, keepaspectratio]{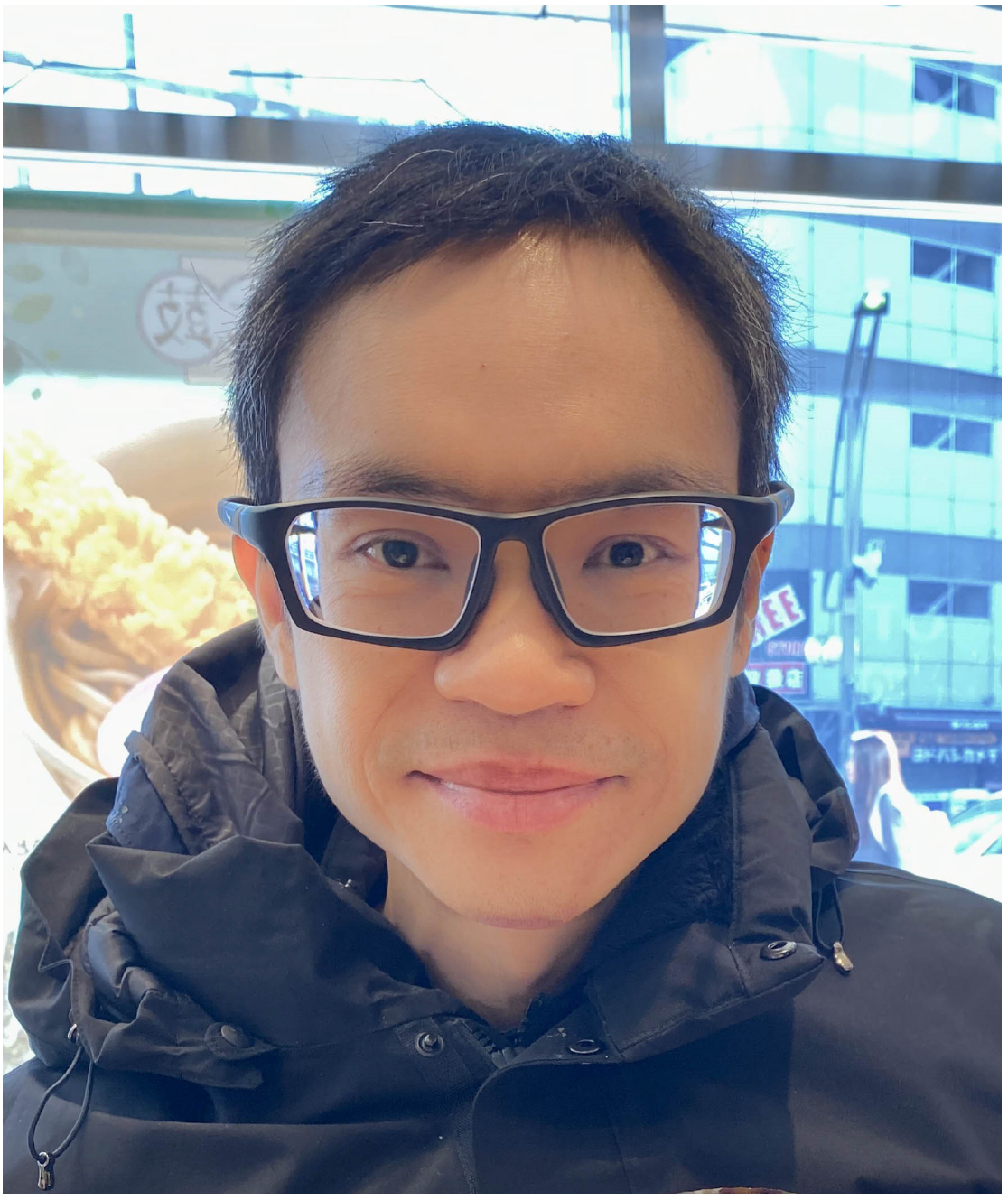}}]{Dusit Niyato} (M'09-SM'15-F'17, IEEE)
	  is a professor in the School of Computer Science and Engineering, at Nanyang Technological University, Singapore. He received B.Eng. from King Mongkut's Institute of Technology Ladkrabang (KMITL), Thailand in 1999 and Ph.D. in Electrical and Computer Engineering from the University of Manitoba, Canada in 2008. His research interests are in the areas of sustainability, edge intelligence, decentralized machine learning, and incentive mechanism design.
	\end{IEEEbiography}

	\begin{IEEEbiography}[{\includegraphics[width=1in, height=1.25in, clip, keepaspectratio]{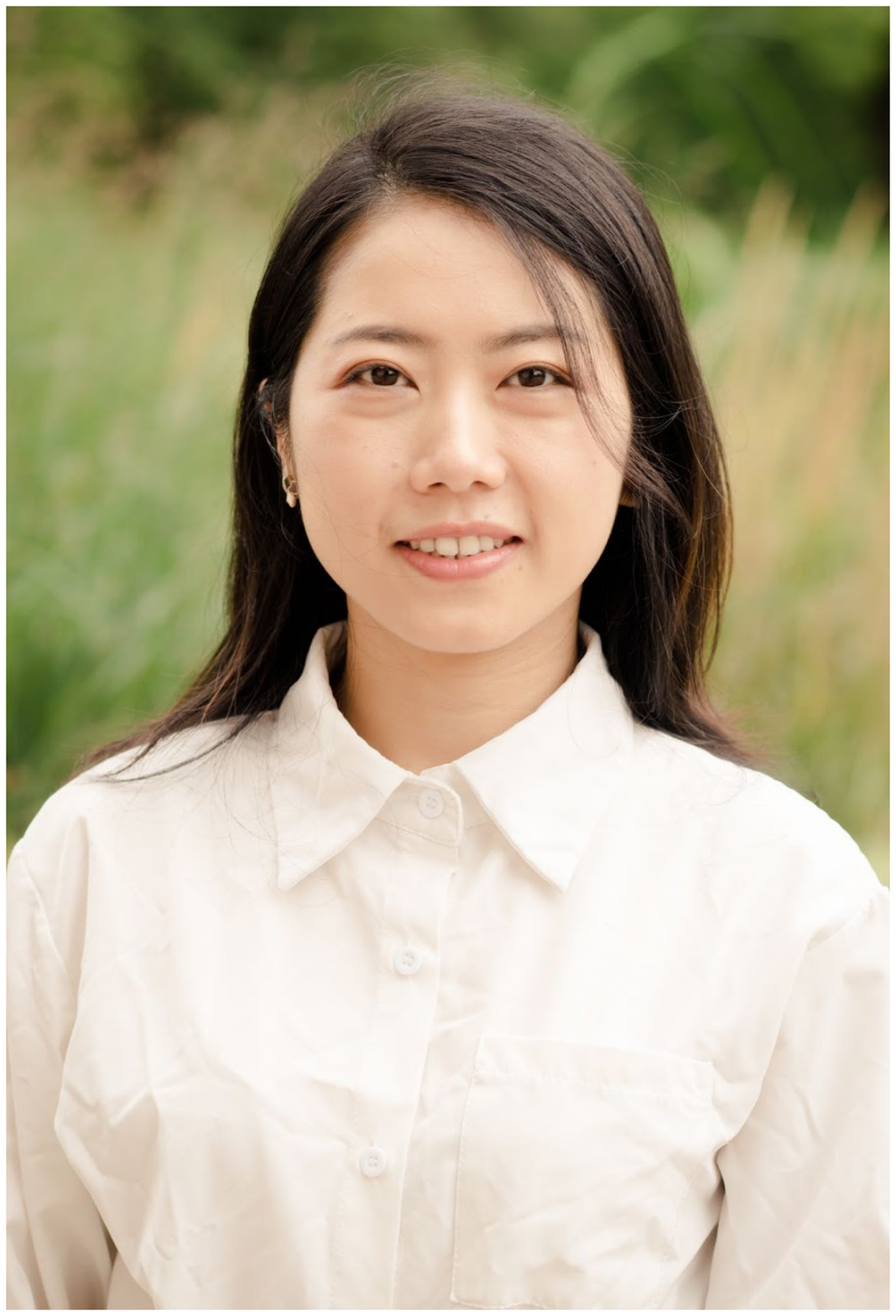}}]{Lan Zhang} (Member, IEEE)
	  received the BE and MS degrees from the University of Electronic Science and Technology of China in 2013 and 2016, respectively, and the PhD degree from the University of Florida in 2020. She has been a tenure-track assistant professor with the Department of Electrical and Computer Engineering at Clemson University since 2024. Before that, she was an assistant professor with the Department of Electrical and Computer Engineering at Michigan Technological University from 2020 to 2023. Her research interests include wireless communications, distributed machine learning, and cybersecurity for various Internet-of-Things applications.
	\end{IEEEbiography}
	\begin{IEEEbiography}[{\includegraphics[width=1in, height=1.25in, clip, keepaspectratio]{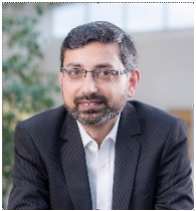}}]{Muhammad Ali Imran} (Fellow, IEEE) 
	received his M.Sc. (Distinction) and Ph.D. degrees from Imperial College London, UK, in 2002 and 2007, respectively. He is a Professor in Communication Systems in the University of Glasgow. He has a global collaborative research network spanning both academia and key industrial players in the field of wireless communications. He has supervised 50+ successful PhD graduates and published over 600 peer-reviewed research papers including more than 100 IEEE Transaction papers. Prof. Imran is a Fellow of IEEE.
	\end{IEEEbiography}

\end{document}